\documentclass[lettersize,journal]{IEEEtran}
\usepackage{amsmath,amssymb,amsfonts}
\usepackage{array}
\usepackage{textcomp}
\usepackage{stfloats}
\usepackage{url}
\usepackage{verbatim}
\usepackage{graphicx}
\def\BibTeX{{\rm B\kern-.05em{\sc i\kern-.025em b}\kern-.08em
    T\kern-.1667em\lower.7ex\hbox{E}\kern-.125emX}}
\usepackage{balance}

\usepackage{textcomp}
\usepackage{hyperref}

\usepackage[ruled,vlined]{algorithm2e}

\usepackage{multirow}
\usepackage{makecell}
\usepackage{subcaption}
\usepackage{booktabs} 

\usepackage{threeparttable}
\usepackage{xcolor}
\usepackage{flushend}

\begin{document}
\title{Multi-Agent Deep Reinforcement Learning for \\Multiple Anesthetics Collaborative Control}

\author{Huijie Li, Yide Yu, Si Shi, Anmin Hu, Jian Huo, Wei Lin, Chaoran Wu, Wuman Luo
\thanks{This work was supported in part by the Macao Polytechnic
University – Research on Representation Learning in Decision
Support for Medical Diagnosis (RP/FCA-11/2022). (Corresponding author: Wuman Luo.)}
\thanks{H. Li, Y. Yu, S. Shi, W. Luo are with Macao Polytechnic University, Macao, China (e-mail: \{huijie.li, yide.yu, si.shi, luowuman\}@mpu.edu.mo)}
\thanks{A. Hu, J. Huo, W. Lin are with Shenzhen United Scheme Technology Co., Ltd, Shenzhen Gd, China (e-mail: \{anmin, huojian, linwei\}@szus.org )}
\thanks{A. Hu, C. Wu are with Department of anesthesia, Shenzhen People's Hospital, Shenzhen Gd, China (e-mail: wu.chaoran@szhospital.com)}}

\markboth{Journal of \LaTeX\ Class Files,~Vol.~18, No.~9, September~2020}%
{How to Use the IEEEtran \LaTeX \ Templates}

\maketitle

\begin{abstract}
Automated control of personalized multiple anesthetics in clinical Total Intravenous Anesthesia (TIVA) is crucial yet challenging. Current systems, including target-controlled infusion (TCI) and closed-loop systems, either rely on relatively static pharmacokinetic/pharmacodynamic (PK/PD) models or focus on single anesthetic control, limiting personalization and collaborative control. 
To address these issues, we propose a novel framework, Value Decomposition Multi-Agent Deep Reinforcement Learning (VD-MADRL). VD-MADRL optimizes the collaboration between two anesthetics propofol (Agent I) and remifentanil (Agent II). And It uses a Markov Game (MG) to identify optimal actions among heterogeneous agents. We employ various value function decomposition methods to resolve the credit allocation problem and enhance collaborative control. We also introduce a multivariate environment model based on random forest (RF) for anesthesia state simulation. Additionally, a data resampling and alignment technique ensures synchronized trajectory data. Our experiments on general and thoracic surgery datasets show that VD-MADRL performs better than human experience. It improves dose precision and keeps anesthesia states stable, providing great clinical value.

\end{abstract}

\begin{IEEEkeywords}
multi-agent deep reinforcement learning, value function decomposition, multiple anesthesia states, personalized anesthesia
\end{IEEEkeywords}

\section{Introduction}
\label{sec:introduction}
Automated control of personalized multiple anesthetics in clinical Total Intravenous Anesthesia (TIVA) is of great significance and remains an urgent problem to be solved at present~\cite{multi4,multi9}. Typically, automated anesthesia control systems can be divided into target-controlled infusion (TCI) systems~\cite{multi1,multi15} and closed-loop systems~\cite{closeloop1,closeloop2,multi11}. TCI systems utilize Pharmacokinetic/Pharmacodynamic (PK/PD) models~\cite{pkpd} to preset and adjust anesthetic dosages to try to achieve the target drug concentrations. 
PK models typically use statistical methods based on population-average data to describe the absorption, distribution and metabolism of anesthetic in the body, thereby predicting drug concentration changes. PD models quantify the pharmacological effects based on the drug concentrations predicted by the PK models.
However, PK/PD models capture individual patient differences based only on static demographic data, and its over-reliance on static fixed-parameter models limits its adaptability to personalized dosage control and unforeseen circumstances. Compared with TCI systems, closed-loop systems can dynamically adjust anesthetic dosages by continuously monitoring physiological signals, thus offering better accommodation to patients' actual conditions. In this paper, we focus on \textbf{P}ersonalized \textbf{M}ultiple \textbf{A}nesthetics \textbf{C}ontrol in a \textbf{C}losed-\textbf{L}oop system (PMAC-CL).

However, PMAC-CL is very challenging due to two main reasons:

\textbf{1) There is a lack of research on the collaborative control of multiple anesthetics.} Most current studies have focused on single anesthetic control~\cite{multi5,multi14,multi3,close_multi_1,close_one_single_1,close_one_single_2,close_two_single_1,close_two_single_2}, which cannot meet clinical anesthesia needs. In clinical TIVA, anesthesiologists usually use multiple anesthetics for a patient to reduce his/her dependence on a single drug and minimize the side effects of anesthetics~\cite{multi6}. 
In 2020, Joosten et al.~\cite{multi8} used three separate Proportional-Integral-Derivative (PID)~\cite{multi5} controllers to automatically control propofol, remifentanil, and ventilation.
However, due to the lack of overall collaboration, they failed to determine the relative contribution of each controller to the overall anesthesia effect. As a result, the risk of overuse of a single anesthetic remains high.

\textbf{2) The research on personalized anesthesia control is still insufficient.} Most current studies~\cite{multi5,multi14,multi3,close_multi_1,close_one_single_1,close_one_single_2} have relied on PK/PD models and Bispectral Index (BIS)~\cite{multi17} to simulate the patient's anesthesia state. This approach has largely limited the personalization degree of automated anesthesia systems. In particular, PK/PD models are mainly based on static demographic data and cannot fully capture the individual behaviours of different patients~\cite{multi5,multi8}. Furthermore, although BIS can provide real-time feedback on cortical activity, it fails to consider other critical aspects of anesthesia depth such as immobility and autonomic responses. In addition to BIS, some researchers have utilized two other indicators: heart rate (HR) and mean arterial blood pressure (MBP)~\cite{close_two_single_1,close_two_single_2}. However, this is not enough compared to the parameters (e.g., respiratory rate (RR) and body temperature (BT)~\cite{multi1}) that need to be considered in practical anesthesia to simulate the anesthetic state.

To address these issues, in this paper, we propose a novel framework called \textbf{V}alue \textbf{D}ecomposition \textbf{M}ulti-\textbf{A}gent \textbf{D}eep \textbf{R}einforcement \textbf{L}earning (VD-MADRL) based on Markov Game (MG) for PMAC-CL. The objective is to effectively explore the collaboration of two anesthetics: the anesthesia propofol (Agent I) and the analgesia remifentanil (Agent II), and to better simulate the anesthesia state based on multiple indicators considered in practical anesthesia. Specifically, in stead of using the commonly used Markov Decision Process (MDP)~\cite{markov}, we propose to abstract the anesthesia process as a MG. Compared to MDP, MG can rapidly identify the optimal actions among heterogeneous agents by abstracting different agents' action spaces into a joint action space. To solve the credit allocation problem~\cite{multiagent} between two anesthetics on the overall anesthesia effect, we use a variety of value function decomposition methods to explore the effects of different collaboration modes between the two anesthetic agents. Moreover, we build a multivariate environment model based on random forest (RF) for multivariate anesthesia state simulation. To effectively learn the internal and external variability of different patients, this model integrates multiple parameters including demographic data , BIS, vital signs data (MBP, BT, HR, RR), PK/PD data and infused dose data.

In addition, we design a data resampling and alignment technique to synchronize trajectory data from different devices. We observed that, in practical anesthesia, the trajectories from different devices usually have different sampling rates, and trajectories from the same device tend to have varied start recording times. These trajectories with misaligned samples will lead to gradient explosion. Our proposed technique can not only effectively avoid the problem of gradient explosion, but also ensure that the trajectory format conforms to the Markov property. To evaluation the performance of our proposed model, we conduct extensive experiments using general surgery and thoracic surgery datasets. The experiment results demonstrate that, compared with human experience, VD-MADRL provides more refined dose adjustments in both online and offline modes, while maintaining multiple anesthesia states more stable at target levels.

Our main contributions are as follows:
\begin{enumerate}
    \item We propose a novel framework called VD-MADRL for PMAC-CL. Besides, we abstract the anesthesia process as a MG, which can explicitly design and optimize collaboration between heterogeneous agents. 
    \item We use different value function decomposition methods to effectively detect the effect of different modes of collaboration between multiple agents. We also develop an environment model based on RF to simulate multiple anesthesia state. Besides, we propose a data resampling and data alignment method to synchronize trajectory data from different devices.
    \item We conduct comprehensive experiments using general surgery and thoracic surgery datasets. Experiment results show that our proposed VD-MADRL outperforms the human experience from the perspectives of dose adjustment and the stability of multiple anesthesia states.
    \item Our model demonstrates substantial clinical value through its exceptional flexibility in the synergistic control of multiple anesthetics. It has also shown a strong correlation between real-time anesthetic dose adjustments and the anesthesia depth index, indicating its effectiveness in optimizing the anesthesia effect.
\end{enumerate}

The rest of this paper is organized as follows: In Section II, we review related work in closed-loop systems and anesthesia state. Section III details the design and implementation of our VD-MADRL model. Section IV presents the experimental results and analysis. Finally, we summarize our paper in Section V. 

\section{Related Work}
\subsection{closed-loop system}
\begin{figure}[t!]
\centering
\includegraphics[width=0.8\columnwidth]{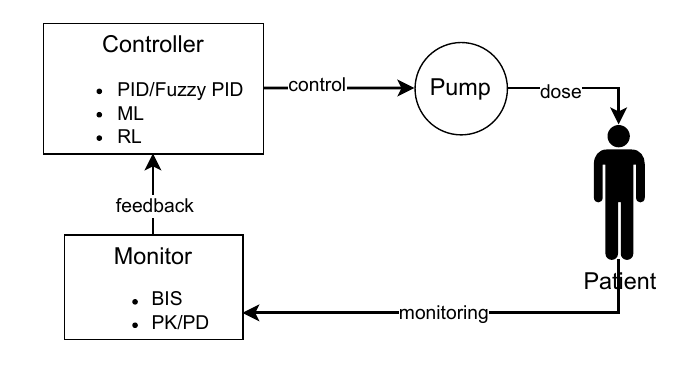}
\caption{Overview of Closed-Loop Systems.}
\label{fig:closedloop}
\end{figure}
Typically, automated anesthesia control systems can be divided into TCI systems~\cite{multi1,multi15} and closed-loop systems~\cite{closeloop1,closeloop2,multi11}. At present, the pure PK/PD-based TCI systems can no longer meet the needs of researchers due to the limitation of the fixed-parameter model, and more attention has been paid to the study of closed-loop systems. As shown in Fig.~\ref{fig:closedloop}, to maintain the patient in an optimal anesthetic state, closed-loop systems dynamically adjust the administration of anesthetics by directly responding to real-time monitored physiological signals. They use different methods to design controllers and incorporate different monitoring metrics to control the controller's decisions.
The design approaches for these controllers can be broadly categorized into three groups. The first group is the traditional parametric and rule-based controllers including PID controller~\cite{multi5,multi8} and fuzzy controller~\cite{multi11}. In detail, Oshin~\cite{multi5} designed two PID controllers, including a linear MPC strategy and a nonlinear strategy, to explore control methods suitable for clinical applications. Mendez~\cite{multi11} design a new fuzzy logic tool using heuristic knowledge provided by clinicians to release the clinician from routine tasks. These methods require empirical manual adjustment of fixed parameters to control the anesthetic dose and do not adequately account for individual patient differences.
The second group is a Machine Learning (ML) based classification controller~\cite{ml_1,ml_2}. Miyaguchi~\cite{ml_1} formulated the decision problem of increasing the flow rate of analgesic remifentanil at each time step as a supervised binary classification problem. These studies also ignore the impact of variability of different indices within and between individuals on infusion control strategies. 

The third group is the most popular reinforcement learning based controller. Reinforcement learning (RL)~\cite{singlerl} techniques have innate advantages for the automated control of anesthesia. RL can interact with the environment and learn to optimize control strategies. In closed-loop system, the agent acts as a controller and adjusts the dose of anesthetic in real time based on the patient's anesthesia state. Current research mainly focuses on single-agent RL for the control of single anesthetic~\cite{multi3,close_multi_1,multi14,close_one_single_1,close_one_single_2,close_two_single_1,close_two_single_2}. Moore's~\cite{multi14,close_one_single_2} first application of reinforcement learning to a closed-loop system to solve a patient-specific control problem improved the accuracy and stability of anesthesia control. Lowery~\cite{close_one_single_1} use actor-critic RL and Yun~\cite{multi3} use hierarchical RL for controlling the flow rate of propofol. In clinical TIVA, anesthesiologists often use multiple anesthetics on patients to reduce the patient's dependence on a single drug and the side effects of anesthetics~\cite{multi6}. Therefore, it is clear that single anesthetic control does not meet clinical anesthesia needs.
\subsection{Anesthesia State}
PK/PD is a statistical model based on demographic data who provides a theoretical framework for understanding and predicting the behavior of anesthetic in the body. Pharmacokinetics (PK) describes the distribution and metabolism of anesthetic, while pharmacodynamics (PD) assesses the effects of anesthesia on the anesthetic concentrations~\cite{multi10}. These PK/PD models were originally applied to the TCI system~\cite{multi1,multi15}. And target-controlled infusion pumps record metrics such as target concentration, plasma concentration, effect-site concentrations, and infusion volume. These metrics were then used to build a PK/PD model to estimate the effect-site anesthetic concentration and to guide dosing decisions. However, PK/PD models are based on static demographic data and cannot fully capture the individual behaviours of different patients.
On the other hand, the BIS~\cite{multi17} is a processed EEG parameter that directly measures a patient's level of consciousness. BIS values range from 0 (complete cortical EEG suppression) to 100 (complete wakefulness), and the target range for general anesthesia is usually 40-60. Because of its ease of use and real-time feedback, BIS has been widely used as an indicator of Depth of Anesthesia (DoA). However, BIS primarily reflects cortical activity and may not fully capture all aspects of DoA, such as immobility or autonomic responses. Moreover, the clinical anesthesiologist will also consider the patient's blood pressure, heart rate, respiratory rate, body temperature and other indicators, and adjust the dose of anesthetics differently for each individual~\cite{multi1}. With these considerations, in our work, we develops a multivariate environment model by combining the PK/PD model, BIS, demographic data, vital signs data and infused dose data to provide a more comprehensive anesthesia state. This approach is consistent with the trend toward more personalized and precise anesthesia management.
\section{VD-MADRL FRAMEWORK}
In this section, we first overview the framework of VD-MADRL. Then, we present the data preprocessing methods. After that, we describe the design of the environment simulator and the implementation process of VD-MADRL, respectively.
\subsection{Overview of Framework}
\label{framework}
\begin{figure}[t!]
\centering
\includegraphics[width=\columnwidth]{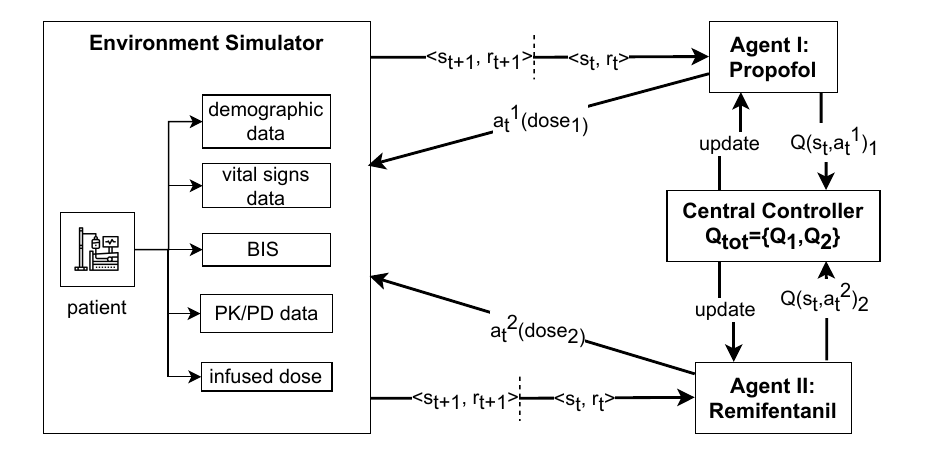}
\caption{VD-MADRL for multiple anesthetics control. VD-MADRL uses two agents to control Propofol and Remifentanil doses, and a central controller decomposes global objectives into local value functions for optimal anesthesia management.}
\label{fig:madrl}
\end{figure}

Our framework consists of two parts, the first part is the multiple anesthesia state indicators environment model used to interact with the agent. The second part is the central controller model based on value fucntion decomposition. Fig.~\ref{fig:madrl} shows the overview of VD-MADRL framework. Agent I controls the dose of Propofol, and agent II controls the dose of Remifentanil. The central controller converts global goals into local goals based on a variety of value function decomposition methods, i.e., the overall joint action value $Q_{tot}$ is decomposed into the independent action values $Q_i$ for updating the policy parameters of each agent.

\begin{figure}[t!]
\centering
\includegraphics[width=\columnwidth]{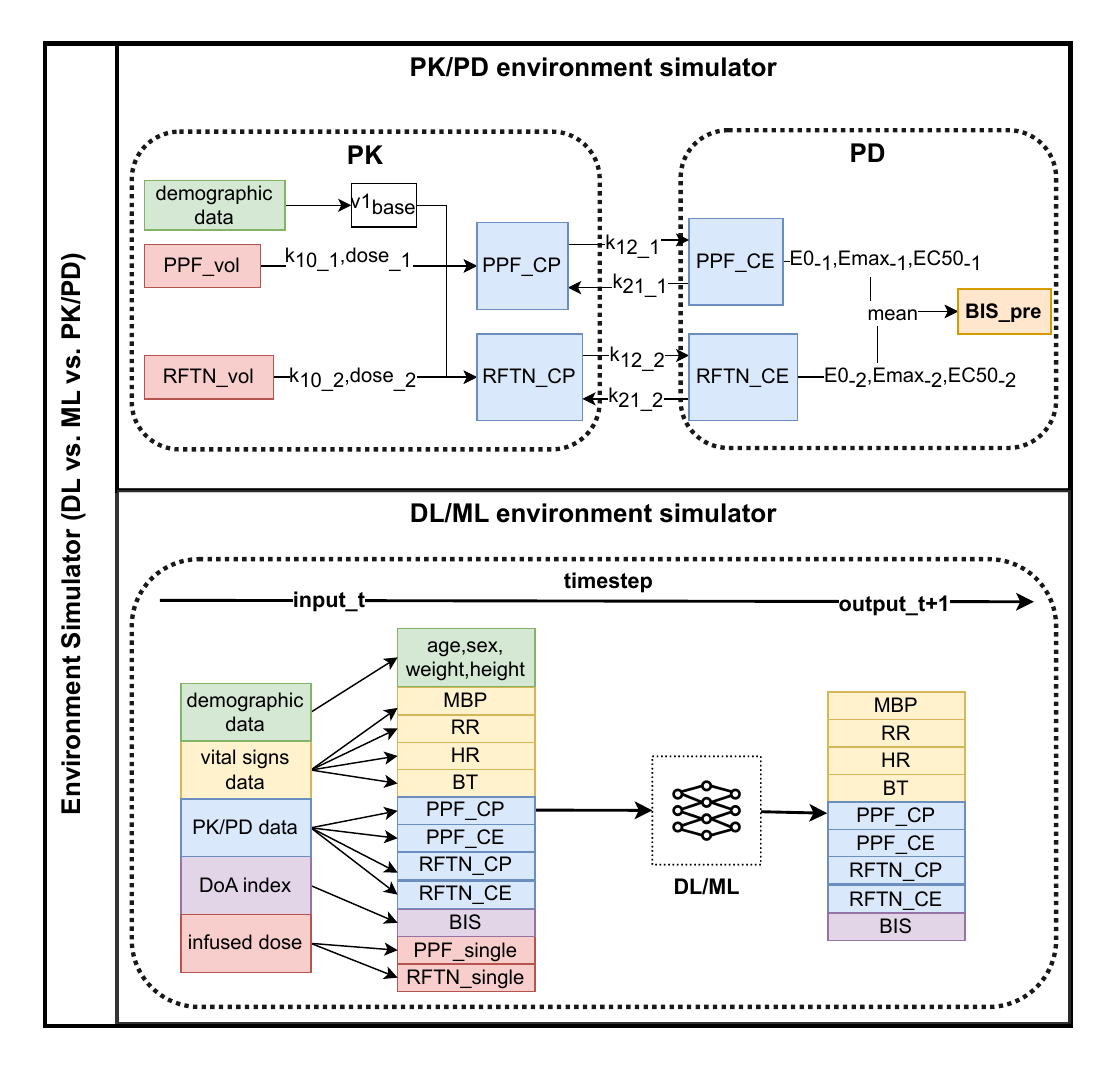}
\caption{Comparison of design environment simulator using various methods. A comparison between traditional PK/PD models and ML/DL-based simulators, highlighting the broader prediction capabilities of ML/DL models for various anesthesia state indicators.}
\label{fig:simulator}
\end{figure}
Fig.~\ref{fig:simulator} compare the traditional PK/PD models, Deep Learning (DL) models and ML models for environment model. PK/PD models typically use infused dose data and demographic data to predict BIS. For the ML and DL models, we integrate 15 anesthesia state indicators at the same time step. These data at time \( t \) serve as inputs to the model. And outputs are at time \( t+1 \) for a total of 9 indicators. Finally, we find that Random Forest (FR)-based is the most effective method for predicting multiple anesthesia state indicators.

\subsection{Data Preprocessing}
\label{datapre}
We performed two preprocessing steps, data resampling and alignment, to synchronise  data from different devices.
The key to the data preprocessing is ensuring that data from different devices are time-aligned for Markov. 
\subsubsection{Data Resampling}
We use data from different devices, each with different recording frequencies. For example, the BIS Vista records data every second, while the Solar 8000M records data every two seconds. To obtain data with the same frequency, we resample the data. Because there is a time delay between anesthetic infusion and BIS value changes, we resample the data at 30-second intervals to better reflect BIS changes. For data recorded every two seconds, we downsample it to 30-second intervals using a 15-second interval method. For data recorded every second, we downsample it to 30-second intervals directly.
\subsubsection{Data Alignment}
\begin{figure}[t!]
\centering
\includegraphics[width=1\linewidth]{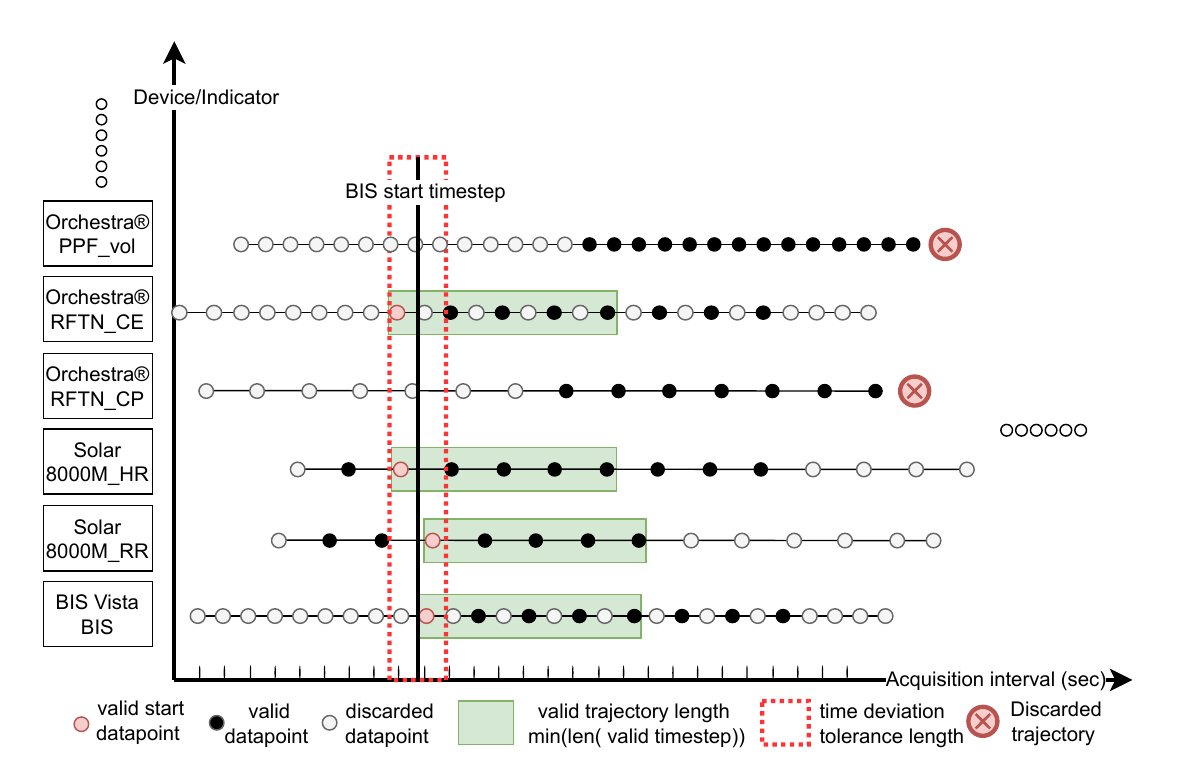}
\caption{Data Alignment. A schematic of the key steps in the data alignment process, including handling invalid data (gray), valid data (black), and aligned valid trajectories (green shaded). Only cases where all anesthesia state indicators are aligned and valid trajectories are retained for further analysis.}
\label{fig:synchronize}
\end{figure}

After resampling, we obtain trajectory data with the same time intervals. However, the data is not aligned because the start times of recordings from different devices are inconsistent. Even within the same device, the start times of different indicators are not consistent. For example, MBP and HR come from the Solar 8000M, but due to different start times, their relationships vary across cases. We also found misaligned data in model training can lead to gradient explosion due to large differences in first-order differences of the same feature across cases. Therefore, we align the data using a series of steps. The data alignment schematic is shown in Fig.~\ref{fig:synchronize}.
Step 1: Data Extraction
We extract trajectory data related to each caseid. Then, we categorize the Orchestra device data by the type of anesthetic. We process data from other devices by indicators separately. We sort each indicator's data frame by case and time, remove rows with zero or negative values before the first non-zero, non-negative value, and keep the first zero row.
Step 2: Valid Start Time Alignment
We group case data by caseid and process each case individually. We use the earliest start time of BIS data as the valid start time step and select case data where all other features' earliest start times are before this timestamp. Considering our 30-second sampling interval, We set a time deviation tolerance length of the valid start time step, that is, If the start times of other features are within ±30 seconds of the initial timestamp, we consider them synchronized. We then truncate data from all features starting from this adjusted valid start time step.
Step 3: Data Truncation
To ensure data length consistency, we select the shortest feature trajectory length and truncate other datasets to this length. Then we get the valid trajectory length of all features.
Step 4: Data Merging
We merge the processed feature trajectories column-wise to form a complete dataset. This dataset contains synchronized data from different devices.
Step 5: Data Filtering
First, we filter caseid based on specific conditions. We select caseid where the minimum time column values of all feature trajectories are less than 300 seconds to ensure synchronized start times. We then filter by time length, excluding records shorter than 120 seconds or longer than 1000 seconds, ensuring anesthesia duration between 1 and 9 hours.

\subsection{Random Forest based Environment Simulator}
\label{sec_random}
To achieve more personalized anesthesia control, we integrate various types of data, including:
\begin{enumerate}
    \item Demographic data (age, gender, weight, height), denoted as \(X_1\).
    \item Vital signs data (MBP, BT, HR, RR), denoted as \(X_2\).
    \item PK/PD data (plasma concentration of propofol (PPF\_CP), effect-site concentration of propofol (PPF\_CE), plasma concentration of remifentanil (RFTN\_CP), effect-site concentration of remifentanil (RFTN\_CE)), denoted as \(X_3\).
    \item Infused dose data (cumulative infused volume of propofol (PPF\_vol), cumulative infused volume of remifentanil (RFTN\_vol)), denoted as \(X_4\).
    \item Anesthesia depth indicator BIS, denoted as \(X_5\).
\end{enumerate}
To accurately simulate the changes in these anesthesia indicators, we divide anesthesia indicator trajectory data into previous and next step series data. Then we construct the feature matrix, which serves as the model's input, as \(X = [X_1, X_{2_t}, X_{3_t}, X_{4_t}, X_{5_t}]\). The target matrix, representing the model's output, is \(Y = [X_{2_{t+1}}, X_{3_{t+1}}, X_{5_{t+1}}]\). 
After that, we concatenate the anesthesia state trajectories of each case and use bootstrapping to randomly select trajectory segments as samples to build each decision tree. This ensures that each tree uses different data subsets, reducing the correlation between trees and improving the model's generalization ability while controlling overfitting. Finally, we use ensemble learning to average the predictions of multiple decision trees to obtain the final prediction. The mathematical representation of the Random Forest model~\cite{random} is as follows:
\begin{equation}
f_{\text{RF}}(\mathbf{X}) = \frac{1}{B} \sum_{b=1}^B T_b(\mathbf{X}; \Theta_b)\label{eq_rf}
\end{equation}
where \(T_b\) represents the \(b\)-th tree, \(\Theta_b\) is the randomly chosen parameters for the \(b\)-th tree, and \(B\) is the total number of trees.

\subsection{MG in VD-MADRL}
\label{sec_markov}
In current research on closed-loop systems, the automation of anesthesia control is often abstracted as a MDP~\cite{markov}. However, when dealing with the combined use of multiple anesthetics, the MDP is evidently insufficient. We abstract the control process of multiple combined anesthetics as a MG~\cite{markovgame}. The MG is represented by the tuple \((N, S, \{A_i\}, P, \{R_i\}, \gamma)\), where \(i \in N\), and \(N\) represents the number of agents, i.e., the types of anesthetics, with \(N > 1\). \(S\) represents the patient's anesthesia state space. \(\{A_i\}\) represents the action space of one type of anesthetic agent \(i\), and the joint action space of multiple anesthetics is \(A = \{A_1\} \times \cdots \times \{A_N\}\). The state transition probability \(P : S \times A \rightarrow P(S)\) indicates the probability of transitioning to state \(s' \in S\) after action \(a \in A\) is applied to state \(s \in S\). Each agent has a corresponding reward function \(R_i : S \times A \times S \rightarrow \mathbb{R}\), which is the real-time feedback reward signal after the action is applied to the environment. The discount factor \(\gamma \in [0, 1]\). 

\paragraph{State}
Based on clinical anesthesiologists' experience and some academic studies \cite{multi13, multi6, multi3}, we select 15 indicators that anesthesiologists are most concerned about during anesthesia surgery as anesthesia states. These include demographic data (age, sex, weight, height), anesthesia depth indicator BIS, vital signs data (MBP, BT, HR, RR), PK/PD data (PPF\_CP, RFTN\_CP, PPF\_CE, RFTN\_CE) and infused dose data(PPF\_vol, RFTN\_vol). These data come from three different monitoring devices and the patient's EHR data. The three different devices are the BIS monitor, a Target-controlled infusion pump (Orchestra), and a patient monitor (Solar8000). Monitor BIS is used to monitor anesthesia depth, Orchestra is used for drug delivery, and Solar8000 is used for real-time monitoring of vital signs.

\paragraph{Action}
We use two anesthetics including propofol and remifentanil, commonly used in clinical anesthesia, as agent I and agent II, respectively. The infusion volumes of propofol 20 mg/mL and remifentanil 20 mg/mL are taken as the actions of the two agents. According to Moore's study \cite{multi14}, high-frequency changes in the BIS should not be used for making dosing decisions. Users can choose to apply a 15-second or 30-second smoothing window to the BIS measurements. Therefore, we set the infusion volumes of propofol and remifentanil at 30-second intervals as the action time step interval.

\begin{figure}[t!]
    \centering
    \begin{subfigure}[t]{0.5\textwidth}
        \includegraphics[width=\columnwidth]{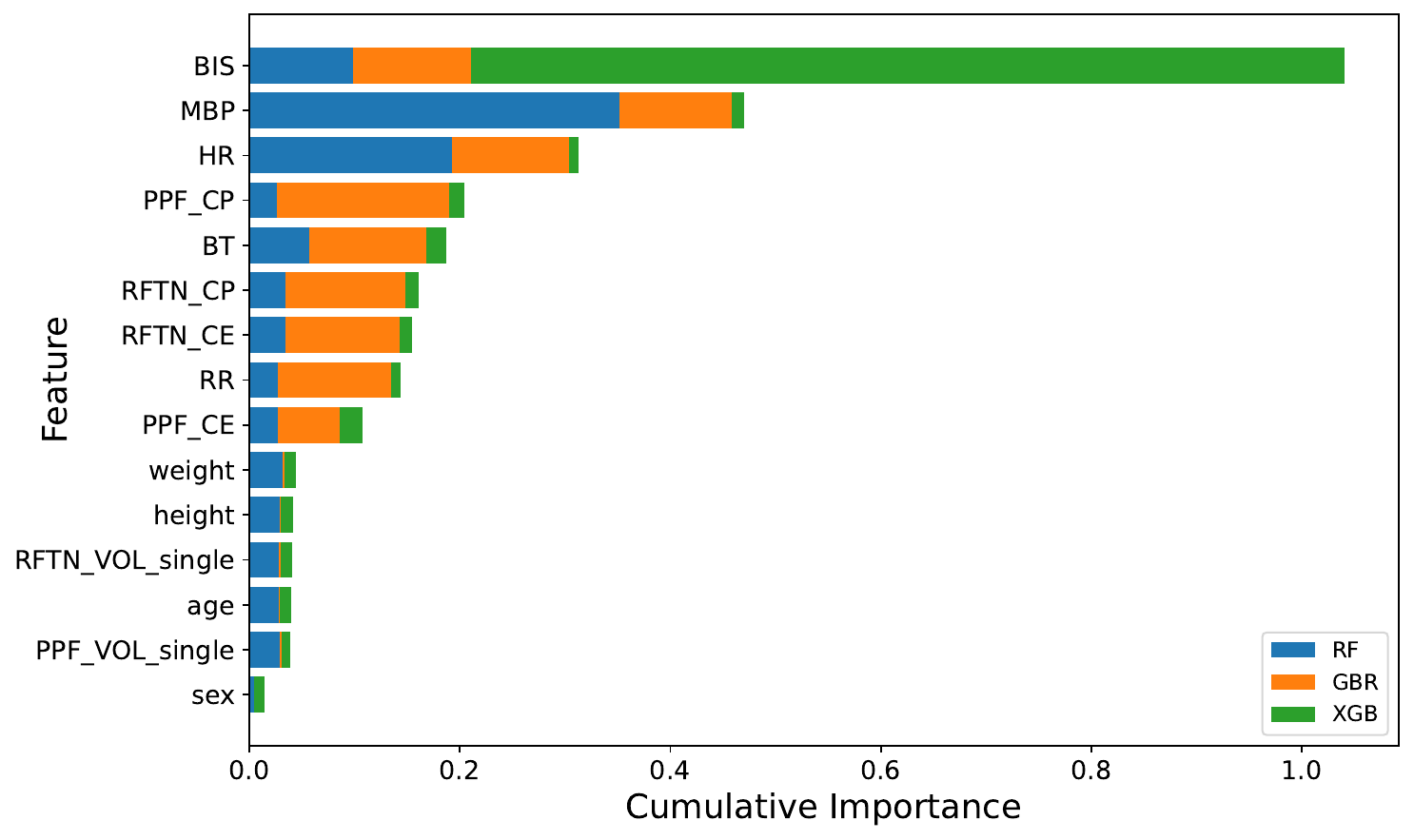}
        \caption{In the General Surgery dataset}
    \end{subfigure}
    \hfill
    \begin{subfigure}[t]{0.5\textwidth}
        \includegraphics[width=\columnwidth]{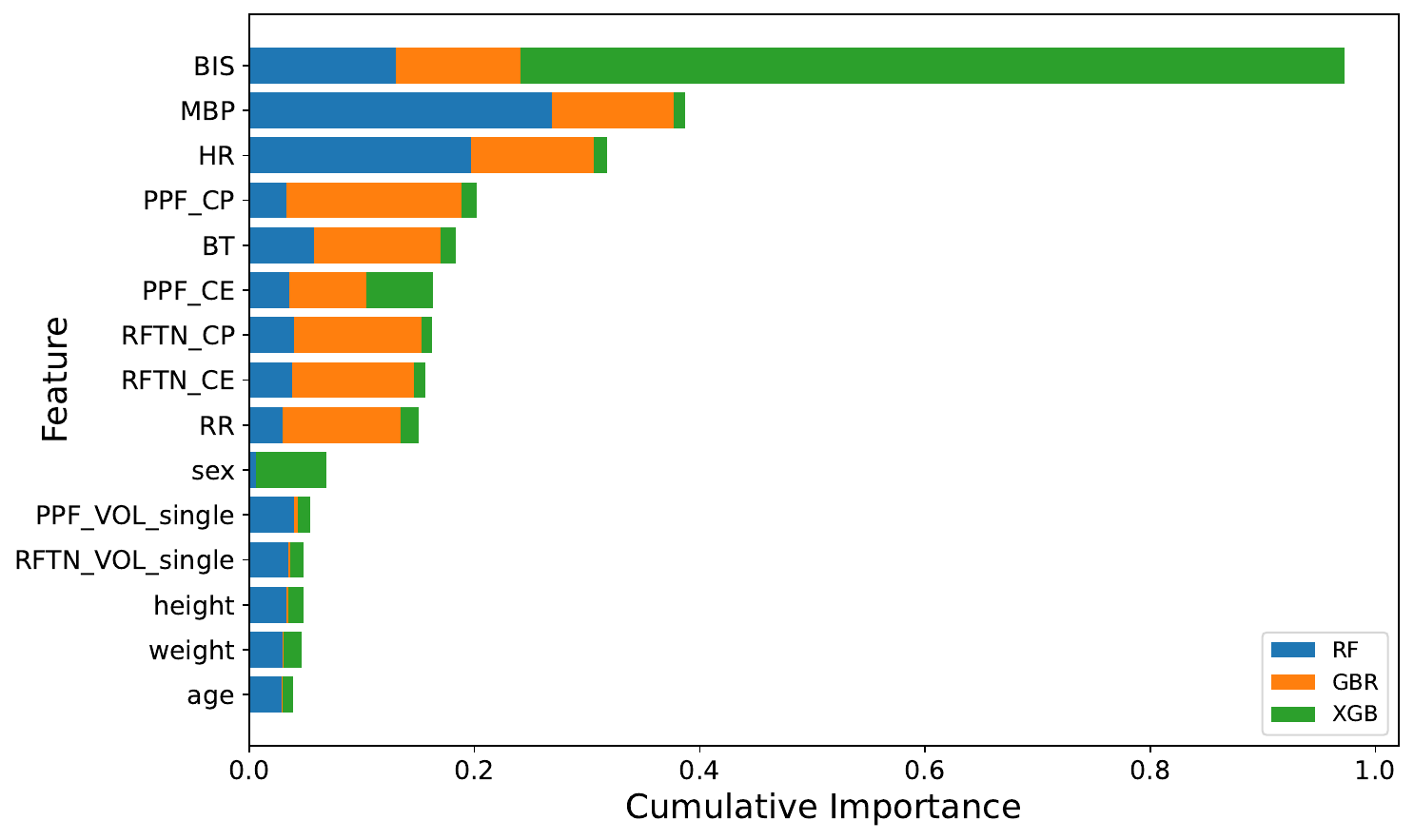}
        \caption{In the Thoracic Surgery dataset}
    \end{subfigure}
    \caption{Cumulative Importance of Anesthesia State Indicators.}
    \label{fig:importance}
\end{figure}

\paragraph{Rewards}
In Fig.~\ref{fig:importance}, we visualized the cumulative importance of various anesthesia state indicators using three different ML models: RF, Gradient Boosting Regression (GBR), and XGBoost (XGB). From the Fig.~\ref{fig:importance}, we observe that BIS has the highest cumulative importance in both datasets. This indicates that BIS most significantly influences the action strategy in anesthesia control. Therefore, we chose BIS as the reward signal of VD-MADRL model to guide the training of our anesthesia strategy. Combined with clinical studies~\cite{multi17}, we set the target BIS value at 50 and the ideal BIS range is 40 to 60. Then we design the reward based on a normal distribution. The reward is maximized when the BIS value is exactly 50. As the BIS value deviates from 50, the reward decreases according to a normal distribution curve, reflecting the degree of deviation from the ideal anesthesia depth. The reward function is defined as follows:
\begin{equation}
    Reward = \exp\left(-\frac{{(bis - \mu)^2}}{{2\sigma^2}}\right)\label{eq_reward}
\end{equation}
where:
\( \mu = 50 \) is the target BIS value,
\( \sigma = 20 \) represents the ideal fluctuation range for the BIS target value.

\subsection{Value Function Decomposition in VD-MADRL}
\label{sec_value}
We compare common value function decomposition methods to evaluate their anesthesia effects in multi-anesthetic automated control. These value function decomposition methods are detailed below.
\paragraph{VDN}
VDN~\cite{vdn} is the most fundamental value function decomposition method. The global value function \( Q_{tot} \) is the sum of all individual agents' \( Q_i \) value functions. We assume that the dosage of each anesthetic has an additive effect on the global anesthesia effect. The global value function in this decomposition method is calculated as follows:
\begin{equation}
    Q_{tot}(\boldsymbol{s}, \boldsymbol{a}) = \sum_{i=1}^{N} Q_i(s, a^i)\label{eq_vdn}
\end{equation}

\paragraph{QMIX}
QMIX~\cite{qmix} uses a mixing network to combine all agents' local Q values to generate a global Q value. This mixing network is subject to a monotonicity constraint, ensuring that the global Q value increases monotonically with any agent's Q value. In multi-anesthetic control, We assume that QMIX can optimize the synergistic effects between the two anesthetics and balance the effects of the two anesthetics on the overall anesthesia effect. When one of the anesthetics deepens the anesthesia effect, the overall anesthesia effect will also deepen without reducing the effect of the other anesthetics on the overall anesthesia effect. The global function expression of QMIX is as follows:
\begin{equation}
    Q_{tot}(\boldsymbol{s}, \boldsymbol{a}) = f(\boldsymbol{s}, Q_1(s, a^1), Q_2(s, a^2), \ldots, Q_N(s, a^N))\label{eq_qmix}
\end{equation}
where \( f \) is a mixing network learned based on the global state \( \boldsymbol{s} \) and is monotonically increasing with each agent's Q value.

\paragraph{CW\_QMIX and OW\_QMIX}
CW\_QMIX and OW\_QMIX~\cite{weightedqmix} are two methods that improve upon QMIX. They dynamically adjust weights based on the achievement of overall goals to encourage exploration of potentially better strategies, especially when current strategies fall short of expectations. The core of CW\_QMIX lies in its center-weight function. Each agent adjusts according to the global goal (maintaining the patient at target anesthesia depth), considering interactions among multiple anesthesia agents when determining the optimal anesthetic delivery strategy. When the system's predicted Q value is below the target Q value, all anesthesia agents are given a higher weight \(1\). In this case, the system believes that more aggressive strategy adjustments are needed to approach the target state. When a specific anesthetic delivery strategy is evaluated as the best action based on the current strategy, it is also given a higher weight. This means that if an adjustment to an anesthetic is deemed the most suitable action in the current environment, it will receive greater weight to drive the implementation of that action. In other cases, if the system's actions do not sufficiently approach the target state, or if an agent's action is not the optimal choice, these actions will be assigned a smaller weight \(\alpha\). This helps to reduce the impact of these actions, thereby allowing room for more effective strategies. The weight control formula for CW\_QMIX is as follows:
\begin{equation}
    w(s, u) = \begin{cases} 
1 & \text{if } y_i > Q^*(s, \tau, u_i^*) \text{ or } u = u^* \\
\alpha & \text{otherwise},
\end{cases}
\end{equation}
where \( u^* = \arg\max_{u} Q_{tot}( \tau, u, s) \), and \( y_i \) is the target computed from the Bellman equation.

OW\_QMIX adopts an optimistic weighting strategy. When the total Q-value estimated by the current policy is lower than the target Q-value, it tends to explore new strategies that may bring higher rewards. Specifically, if the system fails to effectively maintain the patient in an ideal anesthesia state through conventional strategies, OW\_QMIX will increase the weight of actions to encourage the system to try new or less frequently used combinations of anesthetics. The weight control formula for OW\_QMIX is as follows:
\begin{equation}
     w(s, u) = 
\begin{cases} 
1 & \text{if } Q_{tot}(\tau, u, s) < y_i \\
\alpha & \text{otherwise}.
\end{cases}
\end{equation}

\paragraph{QPLEX}
QPLEX~\cite{qplex} extends QMIX by incorporating second-order effects into the joint action-value function. These second-order effects refer to the complex dependencies between agents' actions, which may not be linear or simple additive relationships. QPLEX employs a Dueling network architecture to represent the interactions between different agents. In QPLEX, the total joint action value \( Q_{tot} \) is defined as follows:
\begin{equation}
    Q_{tot} = Q_{duplex} = Q_{mix}(Q_{leaf}(\boldsymbol{s}, \boldsymbol{a}))
\end{equation}
where \( Q_{leaf} \) represents the individual Q-values, \( Q_{mix} \): is a mixing network used to combine the individual Q-values \( Q_{leaf} \) of all agents to produce the total joint action value \( Q_{tot} \).

\paragraph{QTRAN}
The core idea of QTRAN~\cite{qtran} is to learn a transformation function that converts the global action-value function into a decomposable form. This transformation ensures that the optimal joint actions remain consistent between the decomposed and undecomposed value functions. By maintaining consistency and optimization in multi-agent systems' decision-making, QTRAN reduces the likelihood of errors. In the QTRAN framework, the total action-value function \( Q_{tot} \) is defined as follows:
\begin{equation}
    Q_{tot}(\boldsymbol{s}, \boldsymbol{a}) = \sum_{i=1}^{N} Q_i(s_i, a_i) + V(\boldsymbol{s}) - \sum_{i=1}^{N} V_i(s_i)
\end{equation}
Here, \( \boldsymbol{s} \) represents the global state, while \( s_i \) represents the local state of the \( i \)-th agent.
\( \boldsymbol{a} \) represents the joint actions, while \( a_i \) represents the action of the \( i \)-th agent.
\( Q_i(s_i, a_i) \) is the expected reward of the \( i \)-th agent taking action \( a_i \) in local state \( s_i \).
\( V(\boldsymbol{s}) \) is the learned global state value function, providing a value estimate under the global state \( \boldsymbol{s} \).
\( V_i(s_i) \) is the correction term for the \( i \)-th agent, used to align the global value function with the local value functions.

\paragraph{Qatten}
Qatten~\cite{qatten} introduces an attention mechanism to dynamically adjust weights between anesthetic agents. By using the attention mechanism, Qatten can determine the importance of different anesthetic agents' action-value functions. In the Qatten framework, the total action-value function \( Q_{tot} \) is defined as follows:
\begin{equation}
    \begin{aligned}
        Q_{tot}(\boldsymbol{s}, \boldsymbol{a}) &= \text{Attention}(Q_1(s_1, a_1), \ldots, Q_N(s_N, a_N), \boldsymbol{s}) \\
        &= \sum_{i=1}^{N} w_i(\boldsymbol{s}) \cdot Q_i(s_i, a_i)
    \end{aligned}
\end{equation}

Here,
\( \text{Attention} \) is an attention function that dynamically weights the \( Q_i \) values of each agent based on the current global state \( \boldsymbol{s} \) to determine their contribution to \( Q_{tot} \).
\( w_i(\boldsymbol{s}) \) is the weight for the \( i \)-th agent, computed by the attention network. These weights indicate the importance of each agent's decision in the total decision-making process given the current global state. The higher the weight, the greater the influence of the corresponding agent.
\begin{algorithm}[t!]
\SetAlgoLined
\SetKwInOut{Input}{Input}
\SetKwInOut{Output}{Output}
\Input{Clinical anesthesia trajectory data grouped by case ID}
\Output{Trained VD-MADRL model}

\textbf{Online Training} \\
\Begin{
    Initialize network parameters\;
    Set $\epsilon_{\text{initial}} = 0.8$, $\epsilon_{\text{min}} = 0.1$, $\epsilon_{\text{decay}} = 0.01$\;
    \For{each case ID}{
        Select the first record as the initial $S$\;
        \While{not end of trajectory}{
            Select dose using dynamic $\epsilon$-greedy\;
            Pass cumulative dose to simulator\;
            Simulator predicts $S'$\;
            Simulator passes $S',R$ to agents\;
        }
        Perform backpropagation and gradient descent to minimize prediction error\;
        Agents update policy parameters\;
        Periodically update target network\;
    }
}

\textbf{Offline Training} \\
\Begin{
    Store clinical anesthesia Traj data in replay buffer\;
    \While{not converged}{
        Randomly select a case ID trajectory\;
        Extract batch $[S,A,R,S']$\;
        Compute current Q-value based on $S,A$\;
        Compute target Q-value based on target network using Bellman equation\;
        Update network parameters using MSE loss\;
    }
}

\caption{Online and Offline Training modes}
\label{pseudocode}
\end{algorithm}

\subsection{Online/Offline Training for VD-MADRL} 
\label{onoff_mode}
Online training involves dynamic interaction and network parameter updates during the interaction process. This means learning anesthesia strategies through exploration in the simulator. In offline training, all interactions are based on existing clinical anesthesia trajectory data. This mode learns to optimize anesthesia strategies from clinical experts' experiences. For specific training steps see Algorithm~\ref{pseudocode}. 

For the dynamic epsilon-greedy strategy, we initially set a high epsilon value (0.8), meaning there is an 80\% probability of choosing actions randomly at the start of training to ensure sufficient initial exploration probability. In each decision step, the agents decide whether to choose actions randomly (with a probability of \(\epsilon\)) or choose the current estimated optimal action (with a probability of \(1 - \epsilon\)) based on the current epsilon value. This method ensures that the algorithm does not miss potentially useful unexplored paths in the early stages of training and relies more on the model's predictions to make decisions in the later stages and optimize performance. The mathematical expression for dynamically adjusting \(\epsilon\) is:
\begin{equation}
    \epsilon(\text{step}) = \max(\epsilon_\text{min}, \epsilon_\text{initial} - \text{step} \times \epsilon_{\text{decay}})
    \label{eq_epsilon}
\end{equation}
where \text{step} represents the time step of the current case's anesthesia trajectory.

In all the training of multi-agents, we uniformly use the mean square error (MSE) as the loss function, defined as follows:
\begin{equation}
    L(\theta) = \mathbb{E}[(y_t - Q(s, a; \theta))^2]\label{eq_agentloss}
\end{equation}
where \( y_t \) is the target Q value, calculated from the Bellman equation:
\begin{equation}
    y_t = R_{t+1} + \gamma \max_{a'} Q(s', a'; \theta^-)
\end{equation}
here, \( \theta \) denotes the current network parameters and \( \theta^- \) denotes the target network parameters.

\begin{table*}[t!]
\centering
\caption{Comparison of RMSE and $R^2$ score for Different Methods in Predicting Multiple Anesthesia State Indicator Trajectories in General Surgery (Average Values Across n Cases in the Test Set)}
\begin{tabular}{|c|c|c|c|c|c|c|c|c|c|}
\hline
\multicolumn{2}{|c|}{\textbf{Dataset}} & \multicolumn{8}{c|}{\textbf{General surgery}}\\
\hline
\multicolumn{2}{|c|}{\textbf{Method}} & \multicolumn{4}{c|}{\textbf{Machine Learning}} & \multicolumn{3}{c|}{\textbf{Deep Learning}} & {\textbf{PK/PD}} \\
\hline
\textbf{Feature} & \textbf{Metric} & \textbf{RF}& \textbf{GBR}& \textbf{XGboost}& \textbf{SVR}& \textbf{GRU}& \textbf{Transformer}& \textbf{GRU\_Trans}& \textbf{PK/PD}\\
\hline 
\multirow{2}{*}{\textbf{BIS}} &\textit{RMSE}& \textbf{0.0045}
 &0.0163&0.0157&0.0371& 0.0990 & 0.0844& 0.1094 & 0.0331 \\
\cline{2-2}
& \textit{$R^2$}\_\textit{score}& \textbf{0.8964}& 0.63320
 & 0.6572& 0 &  -& -&- &-  \\
\hline
\multirow{2}{*}{\textbf{MBP}} &\textit{RMSE}&\textbf{0.0106}&0.0394&0.0379&0.1346& 0.0913& 0.0869& 0.0932& -\\
\cline{2-2}
& \textit{$R^2$}\_\textit{score} &\textbf{0.8317}&0.5070&0.5398&-7
 &-& -&- &- \\
\hline
\multirow{2}{*}{\textbf{HR}} &\textit{RMSE}&\textbf{0.0038}&0.0136&0.0132&0.0593
 & 0.0807& 0.0715& 0.0678& - \\
\cline{2-2}
& \textit{$R^2$}\_\textit{score} &\textbf{0.6921}
 &0.6478
 &0.6661
 &-12
 & -& -&- &- \\
\hline
\multirow{2}{*}{\textbf{PPF\_CP}} &\textit{RMSE}&\textbf{0.0002}
 &0.0005
 &0.0005
 &0.0095
 & 0.1228& 0.1322 & 0.1701& 0.0094 \\
\cline{2-2}
& \textit{$R^2$}\_\textit{score} &0.8717
 &0.8862
 &\textbf{0.9089}
 &-51
 & -& -&- &- \\
\hline
\multirow{2}{*}{\textbf{BT}} &\textit{RMSE}&\textbf{0.0007}
 &0.0017
 &0.0015
 & 0.0343
& 0.1005& 0.0368 &0.1275 & - \\
\cline{2-2}
& \textit{$R^2$}\_\textit{score} &0.6378
 &0.9666
 & \textbf{0.9719}
&-125
 & -& -&- &-  \\
\hline
\multirow{2}{*}{\textbf{RFTN\_CP}} &\textit{RMSE}&\textbf{0.0004}
 &0.0013
 &0.0013
 &0.0231
 &  0.1336& 0.0601& 0.1759& 0.0206\\
\cline{2-2}
& \textit{$R^2$}\_\textit{score} &\textbf{0.9171}
 &0.7954
 &0.8042
 & -102
& -& -&- &-  \\
\hline
\multirow{2}{*}{\textbf{RFTN\_CE}} &\textit{RMSE}&0.0009&0.0003
 &\textbf{0.0002}
 &0.0078
 & 0.1309&  0.1179& 0.1706& 0.0022 \\
\cline{2-2}
& \textit{$R^2$}\_\textit{score} & 0.9393
&0.9907
 &\textbf{0.9912}
 & -17
&-& -&- &-  \\
\hline
\multirow{2}{*}{\textbf{RR}} &\textit{RMSE}& \textbf{0.0011}
&0.0034
 &0.0030
 &0.0648
 & 0.1210& 0.1506& 0.1577& - \\
\cline{2-2}
& \textit{$R^2$}\_\textit{score} &\textbf{0.7778}
 &0.6163
 &0.6925
 &-244
 &-& -&- &-  \\
\hline
\multirow{2}{*}{\textbf{PPF\_CE}} &\textit{RMSE}& \textbf{0.0002}
&\textbf{0.0002}
 &\textbf{0.0002}
 &0.0055
 & 0.1187& 0.0420& 0.1698&  0.0007 \\
\cline{2-2}
& \textit{$R^2$}\_\textit{score} &0.8899
 &0.9799
 &\textbf{0.9826}
 &-18
 & -& -&- &- \\
\hline
\multirow{2}{*}{\textbf{Total\_Feature}} &\textit{RMSE}& \textbf{0.0091}& 0.0162& 0.0156& 0.0579& 0.1129& 0.0955& 0.1436& 0.0186 \\
\cline{2-2}
& \textit{$R^2$}\_\textit{score} & \textbf{0.8281}& 0.7803& 0.8015& -64& -& -&- &-  \\
\hline
\end{tabular}
\label{table_mse_general}
\end{table*}

\section{Experiment}
In this section, we empirically study the performance of the proposed VD-MADRL. First, we describe the experimental settings, including datasets, baseline methods, and evaluation metrics. Then, we analyze the simulation effect of the environment simulator and the anesthesia effect of our VD-MADRL framework, respectively. Finally, we select four best-performing models for each of the combinations of the two datasets and the two training modes, and compare each of them against the human experience trajectory data.

\subsection{Experiment Settings}
\subsubsection{Datasets}
We create two datasets, i.e., general surgery dataset and thoracic surgery dataset, from the public VitalDB~\cite{vitaldb}. VitalDB is renowned for its high-resolution perioperative patient data. After data preprocessing of VitalDB, we obtained a general surgery set containing 550 cases and a thoracic surgery set containing 459 cases. Each case consists of 11 anesthesia trajectory records, and the duration of each trajectory ranges from 1 to 9 hours. We further split each dataset into training set and test set in a 4:1 ratio.

\subsubsection{Baselines}
We evaluate the performance of the environment model and the agent model separately. We select 7 baselines for the environment model including one traditional PK/PD, three ML models (i.e., GBR, XGB and SVR), and three DL models (i.e., GRU, Transformer and GRU\_Trans). 

\begin{itemize}
    \item PK/PD~\cite{multi_twocompartment}: Use a two-compartment model to describe the dynamic behavior of anesthetic in human body and their pharmacological effects.
    \item GBR~\cite{gbr}: Tree-based ensemble method for fitting residuals. 
    \item XGB~\cite{xgboost}: Ensemble learning tree-based method optimized from the gradient boosting algorithm.
    \item SVR~\cite{svr}: Regression method based on support vector machines.
    \item GRU~\cite{gru}: Recurrent neural network (RNN) architecture designed to capture dependencies in sequential data. 
    \item Transformer~\cite{transformer}: Rely on self-attention mechanisms to process and encode sequential data.
    \item GRU\_Trans: GRU and Transformer hybrid architecture.
\end{itemize}    

As we discussed previously, there is a lack of research on the control of two anesthetics using multi-agent reinforcement learning. Additionally, in practical anesthesia, it is a challenging task for clinical anesthesiologists to coordinate the control two anesthetics~\cite{anesthesiologist}. We therefore employ clinical anesthesiologists' expertise (human experience) as a baseline for our agent model.

\subsubsection{Evaluation Metrics}
For the prediction performance of the environment model, we use the square root of the mean squared error (RMSE) and the coefficient of determination $R^2$ score~\cite{entropy} as metrics. To evaluate the anesthetic effect of the agent's policy, we use the cumulative reward (CR) of the complete anesthesia trajectory for each case and the Median Performance Error (MDPE)~\cite{mdpemdape}.
MDPE quantifies the median deviation of the model-generated anesthesia effect from the target BIS value. It is calculated as:
\begin{equation}
    MDPE = \text{median} \left( \frac{BIS - BIS_{target}}{BIS_{target}} \times 100 \right)
\end{equation}

\begin{table*}[t!]
\centering
\caption{Comparison of RMSE and ${R}^2$ score for Different Methods in Predicting Multiple Anesthesia State Indicator Trajectories in Thoracic Surgery (Average Values Across n Cases in the Test Set)}
\begin{tabular}{|c|c|c|c|c|c|c|c|c|c|}
\hline
\multicolumn{2}{|c|}{\textbf{Dataset}} & \multicolumn{8}{c|}{\textbf{Thoracic surgery}}\\
\hline
\multicolumn{2}{|c|}{\textbf{Method}} & \multicolumn{4}{c|}{\textbf{Machine Learning}} & \multicolumn{3}{c|}{\textbf{Deep Learning}} & {\textbf{PK/PD}} \\
\hline
\textbf{Feature} & \textbf{Metric} & \textbf{RF}& \textbf{GBR}& \textbf{XGboost}& \textbf{SVR}& \textbf{GRU}& \textbf{Transformer}& \textbf{GRU\_Trans}& \textbf{PK/PD}\\
\hline 
\multirow{2}{*}{\textbf{BIS}} &\textit{RMSE}&\textbf{0.0047}
 &0.0194
 &0.0184
 &0.0428
 & 0.0810& 0.1860&0.2596 &0.0386  \\
\cline{2-2}
& \textit{$R^2$}\_\textit{score}&\textbf{0.8902}
 &0.5797
 &0.6235
 &0
 & -& -&- &-  \\
\hline
\multirow{2}{*}{\textbf{MBP}} &\textit{RMSE}&\textbf{0.0078}
 &0.0371
 &0.0356
 &0.1404
 & 0.0895& 0.1013&0.1945 & -\\
\cline{2-2}
& \textit{$R^2$}\_\textit{score} &\textbf{0.8931}
 &0.5547
 &0.5890
 &-7
 & -& -&- &-  \\
\hline
\multirow{2}{*}{\textbf{HR}} &\textit{RMSE}&\textbf{0.0001}
 &0.0230
 &0.0219
 &0.0607
 & 0.0727&0.1452 &0.2134 & - \\
\cline{2-2}
& \textit{$R^2$}\_\textit{score} &\textbf{0.8526}
 &0.3368
 &0.3753
 &-6
 & -& -&- &-  \\
\hline
\multirow{2}{*}{\textbf{PPF\_CP}} &\textit{RMSE}&\textbf{0.0003}
 &{0.0004}
 & {0.0004}
&0.0158
 & 0.1252& 0.2473& 0.3281& 0.0092\\
\cline{2-2}
& \textit{$R^2$}\_\textit{score} &\textbf{0.9086}
 &0.8177
 &0.8611
 &-91650
 & -& -&- &-  \\
\hline
\multirow{2}{*}{\textbf{BT}} &\textit{RMSE}&\textbf{0.0005}
 &0.0013
 &0.0012
 &0.0462
 & 0.0882& 0.0268&0.2650 & - \\
\cline{2-2}
& \textit{$R^2$}\_\textit{score} &0
 &\textbf{0.9616}
 &0.9330
 &-7338
 & -& -&- &-  \\
\hline
\multirow{2}{*}{\textbf{RFTN\_CP}} &\textit{RMSE}&\textbf{0.0004}
 &0.0012
 &0.0011
 &0.0195
 & 0.1240&  0.1616& 0.3289& 0.0111 \\
\cline{2-2}
& \textit{$R^2$}\_\textit{score} &\textbf{0.9189}
 &0.8095
 &0.7976
 &-78
 & -& -&- &-  \\
\hline
\multirow{2}{*}{\textbf{RFTN\_CE}} &\textit{RMSE}&0.0003
 &\textbf{0.0001}
 &\textbf{0.0001}
 &0.0079
 & 0.1253& 0.2161& 0.3286&0.0022  \\
\cline{2-2}
& \textit{$R^2$}\_\textit{score} &0.9343
 &0.9922
 &\textbf{0.9944}
 &-34
 & -& -&- &-  \\
\hline
\multirow{2}{*}{\textbf{RR}} &\textit{RMSE}&\textbf{0.0015}
&0.0053
 &0.0047
 &0.0432
 &  0.1081& 0.2383&0.3066 & - \\
\cline{2-2}
& \textit{$R^2$}\_\textit{score} &\textbf{0.8220}
 &0.5126
 &0.6045
 &-42
 & -& -&- &- \\
\hline
\multirow{2}{*}{\textbf{PPF\_CE}} &\textit{RMSE}&0.0002
 &\textbf{0.0001}
 &\textbf{0.0001}
 &0.0035
 &0.1253 & 0.1555&0.3269 &  0.0009  \\
\cline{2-2}
& \textit{$R^2$}\_\textit{score} & 0.9220
&\textbf{0.9745}
 &0.9725
 &-6212
 & -& -&- &-  \\
\hline
\multirow{2}{*}{\textbf{Total\_Feature}} &\textit{RMSE}& \textbf{0.0078}& 0.0175& 0.0169& 0.0586& 0.1069& 0.1775&0.2596 & 0.0180 \\
\cline{2-2}
&\textit{$R^2$}\_\textit{score} & 0.7377& 0.7265& \textbf{0.7500}& -11708& -& -&- &-  \\
\hline
\end{tabular}
\label{table_mse_thoracic}
\end{table*}

\begin{figure*}[t!]
    \centering
    \begin{subfigure}[t]{0.23\textwidth}
        \includegraphics[width=\textwidth]{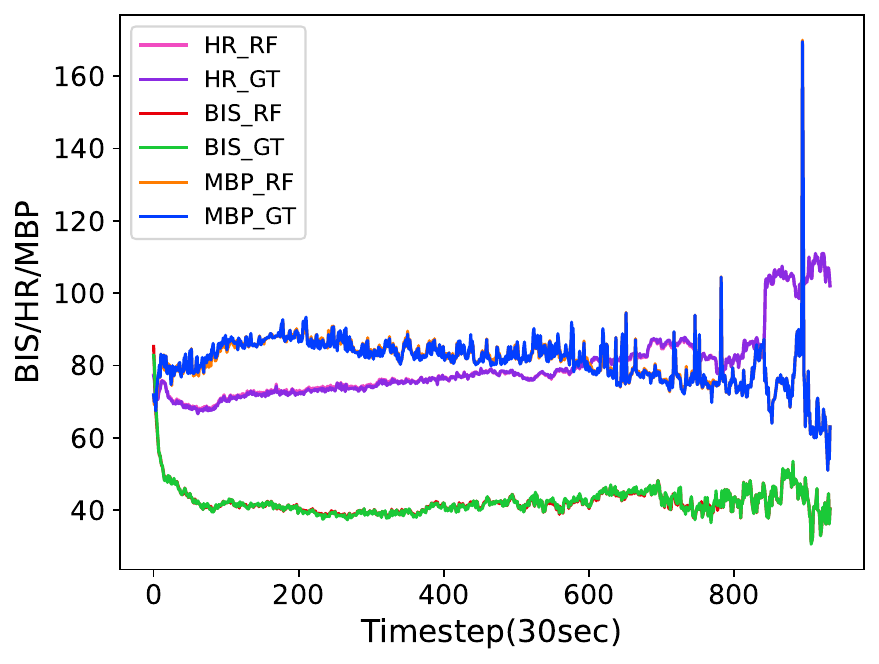}
        \caption{BIS/HR/MBP trajectories}
        \label{fig:SG_mse}
    \end{subfigure}
    \hfill
    \begin{subfigure}[t]{0.23\textwidth}
        \includegraphics[width=\textwidth]{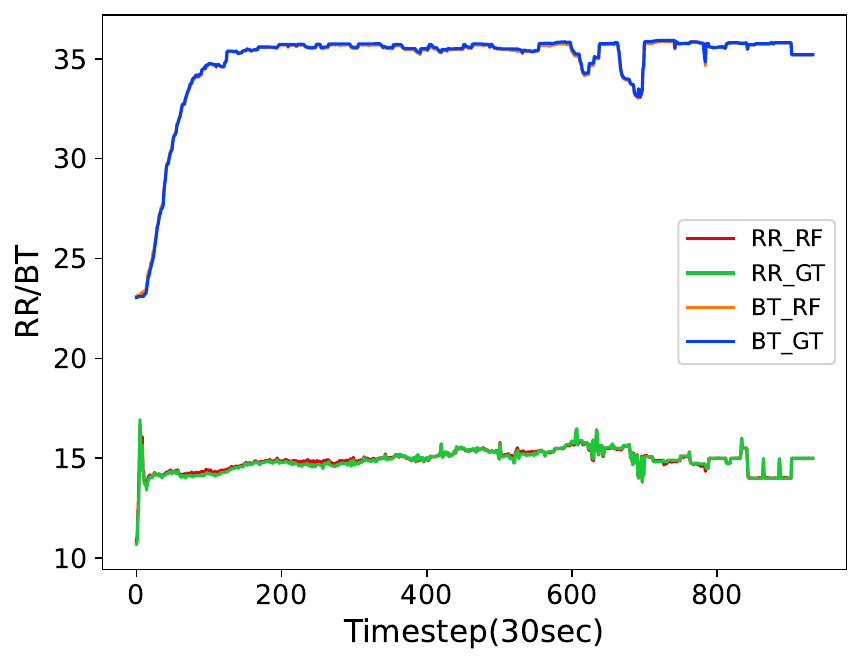}
        \caption{RR/BT trajectories}
        \label{fig:bis}
    \end{subfigure}
    \hfill
    \begin{subfigure}[t]{0.23\textwidth}
        \includegraphics[width=\textwidth]{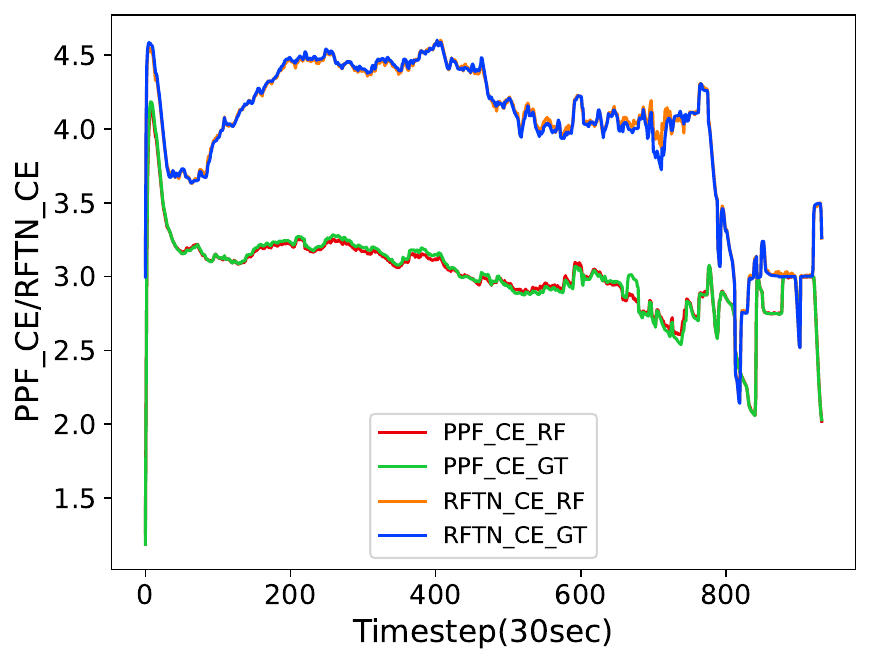}
        \caption{CE trajectories}
        \label{fig:gru}
    \end{subfigure}
    \hfill
    \begin{subfigure}[t]{0.23\textwidth}
        \includegraphics[width=\textwidth]{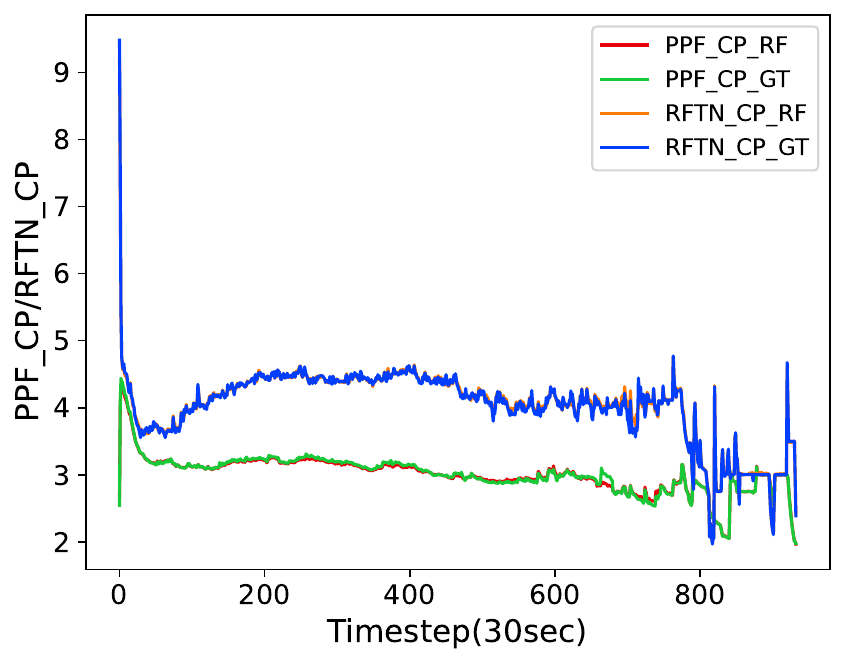}
        \caption{CP trajectories}
        \label{fig:rr}
    \end{subfigure}
    \caption{Comparison of RF Predictions and Actual Anesthesia State Trajectories (GT) in General Surgery Dataset.}
    \label{fig:ml_general}
\end{figure*}

\begin{figure*}[t!]
    \centering
    \begin{subfigure}[t]{0.23\textwidth}
        \includegraphics[width=\linewidth]{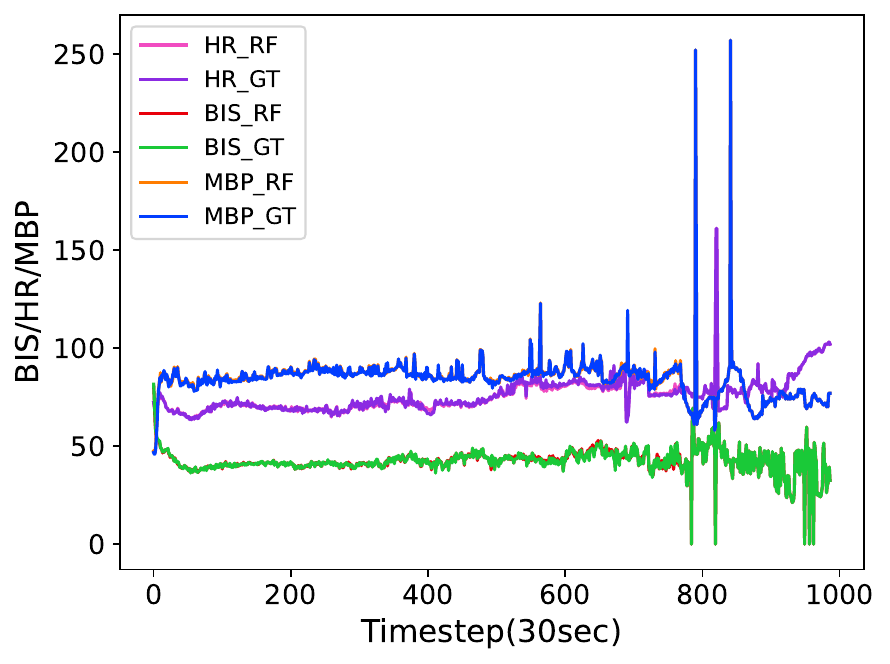}
        \caption{BIS/HR/MBP trajectories}
        \label{fig:SG_mse}
    \end{subfigure}
    \hfill
    \begin{subfigure}[t]{0.23\textwidth}
        \includegraphics[width=\linewidth]{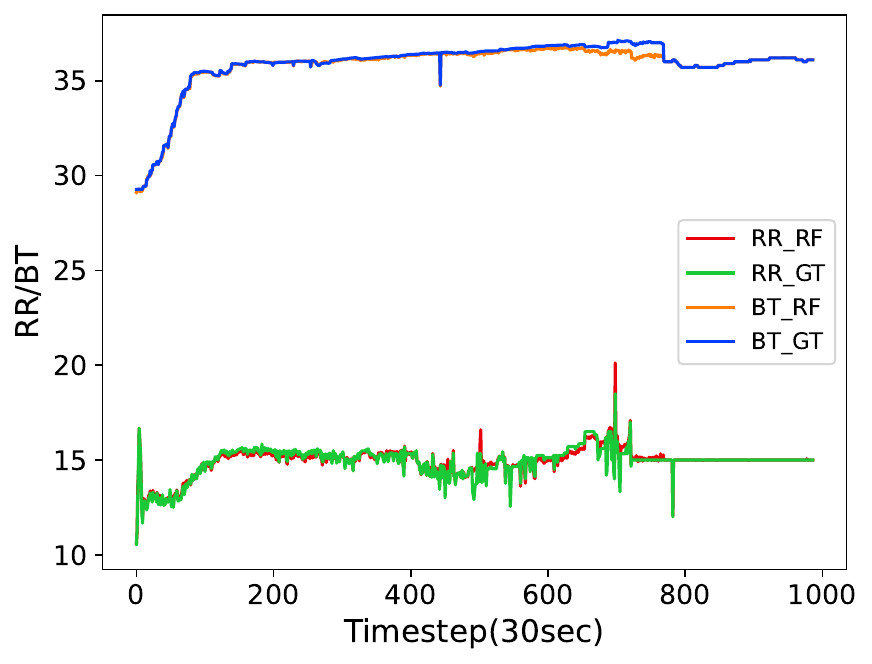}
        \caption{RR/BT trajectories}
        \label{fig:bis}
    \end{subfigure}
    \hfill
    \begin{subfigure}[t]{0.23\textwidth}
        \includegraphics[width=\linewidth]{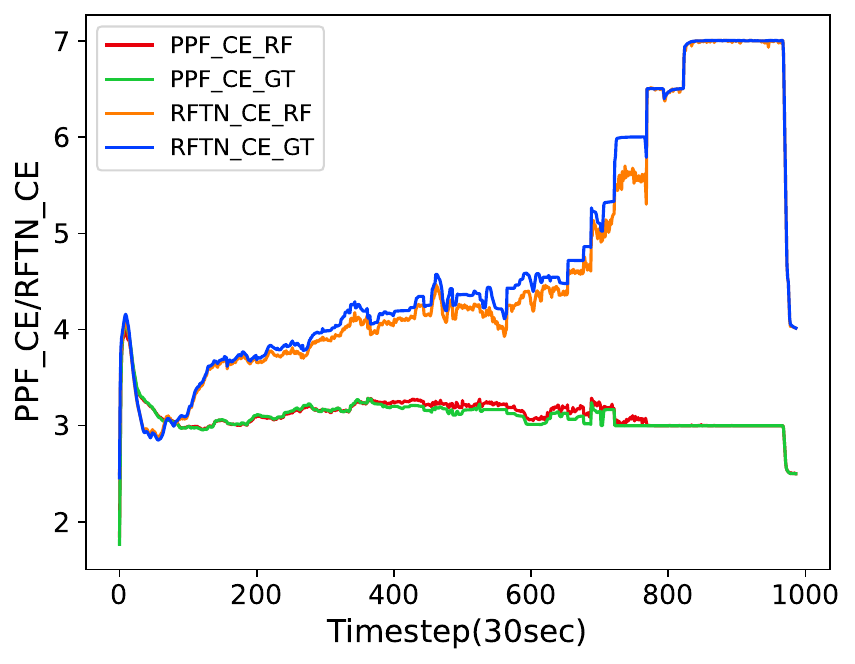}
        \caption{CE trajectories}
        \label{fig:gru}
    \end{subfigure}
    \hfill
    \begin{subfigure}[t]{0.23\textwidth}
        \includegraphics[width=\linewidth]{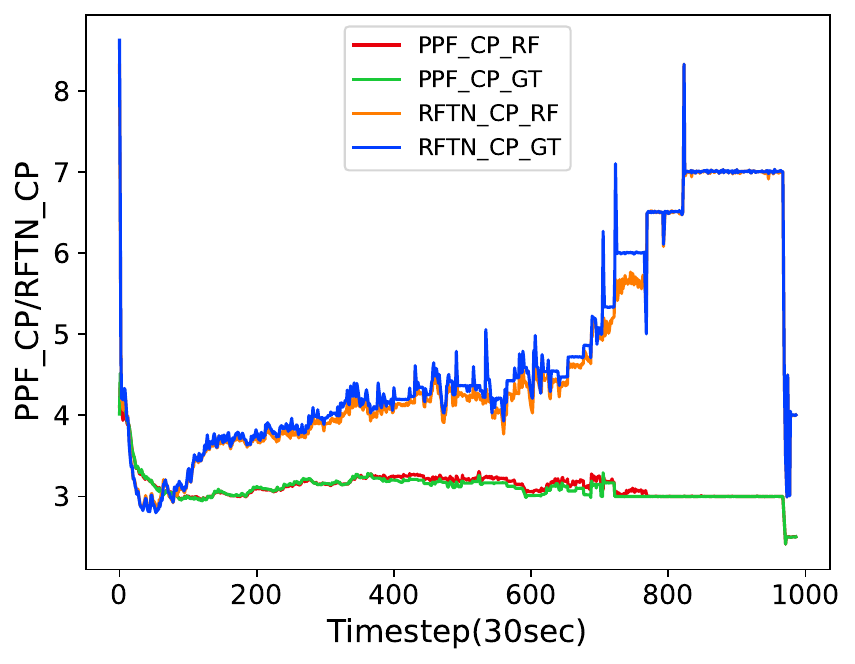}
        \caption{CP trajectories}
        \label{fig:rr}
    \end{subfigure}
    \caption{Comparison of RF Predictions and Actual Anesthesia State Trajectories (GT) in Thoracic Surgery Dataset.}
    \label{fig:ml_thoracic}
\end{figure*}

\subsection{Simulation Effect Analysis of Environment Simulator}
We compare our proposed RF environment simulator with the 7 baselines in terms of RMSE and $R^2$ score using the two datasets. The comparison results are recorded in Table~\ref{table_mse_general} and Table~\ref{table_mse_thoracic}, respectively. Moreover, we compare the RF predictions and the ground truth data using the two datasets, and visualize the comparison results in Fig.~\ref{fig:ml_general} and Fig.~\ref{fig:ml_thoracic}, respectively.

Table~\ref{table_mse_general} records the comparison results for the General surgery dataset, and Table~\ref{table_mse_thoracic} records the results for the Thoracic surgery dataset. As shown in Table~\ref{table_mse_general} and Table~\ref{table_mse_thoracic}, RF consistently achieves the lowest RMSE and highest $R^2$ score for nearly all anesthesia state indicators in both datasets, which demonstrates RF's superior predictive accuracy and high stability. We can also see that all the best results (RMSE and $R^2$) are from the machine learning models. This is because deep learning models highly rely on the refined adjustment of layer parameters and have a high demand for the size and distribution of the datasets. Besides, PK/PD models have poor personalized control for varying individual patient statuses. Although GBR and XGboost achieve the best performance for a few indicators, they cannot consistently perform well compared with RF for most of the indicators on both datasets. 

However, it is worth noting that in the Thoracic surgery dataset (Table~\ref{table_mse_thoracic}), although RF performs well for the majority of indicators, its overall $R^2$ score (0.7377 for Total\_Feature) is slightly lower than that of XGboost ($R^2 = 0.7500$). Moreover, GBR and XGboost achieve the highest predictive accuracy for pharmacodynamic indicators (RFTN\_CE and PPF\_CE), in both datasets. This suggests that the RF model, though robust, might struggle with capturing specific feature dependencies that are more effectively handled by ensemble models like GBR and XGboost. Additionally, RF’s relatively lower $R^2$ score for the BT indicator in both datasets is notable. In the General surgery dataset (Table~\ref{table_mse_general}), RF achieves an $R^2$ score of 0.6378 for BT, while in the Thoracic surgery dataset (Table~\ref{table_mse_thoracic}), its performance is even lower, with an $R^2$ score of 0. This could be because in the context of body temperature, rare events such as sudden hypothermia or hyperthermia caused by specific surgical interventions or patient reactions might be underrepresented in the dataset. Since RF uses averaging across decision trees, this generally makes it robust to noise but also less sensitive to outliers or rare events, which may explain its slightly lower overall $R^2$ score in the Thoracic surgery dataset.

Nevertheless, given that the pharmacodynamic indicators (RFTN\_CE and PPF\_CE) and BT are of relatively lower importance in the overall anesthesia state prediction, as shown in Fig.~\ref{fig:importance}. And considering the fact that for the most important BIS indicator, RF's predictive power is much higher than all baselines in both datasets. So, it can be concluded that RF continues to be a strong and reliable model for the majority of the more important anesthesia state indicators across both datasets.

Fig.~\ref{fig:ml_general} and Fig.~\ref{fig:ml_thoracic} illustrate the predictive performance of our proposed RF environment simulator compared to the actual anesthesia state trajectories (GT) across two datasets: the General Surgery dataset (Fig.~\ref{fig:ml_general}) and the Thoracic Surgery dataset (Fig.~\ref{fig:ml_thoracic}). We integrate different indicator trajectories with similar scale into one subfigure, facilitating an intuitive comparison between our model's predictions and actual values. The x-axis represents the entire anesthesia process in timesteps (measured in 30-second intervals), spanning from the beginning to the end of the surgical procedure. The y-axis represents the average value of each trajectory indicator in the test set, including BIS, HR, MBP, RR, BT, PPF\_CE, PPF\_CP, RFTN\_CE and RFTN\_CP. Specifically, in Fig.~\ref{fig:ml_general}, sub-figure (a) displays BIS, HR, and MBP trajectories; sub-figure (b) shows RR and BT trajectories; sub-figure (c) presents PPF\_CE and RFTN\_CE trajectories; and sub-figure (d) illustrates PPF\_CP and RFTN\_CP trajectories. Similarly, these sub-figures have the same trajectories shown in Fig.~\ref{fig:ml_thoracic}. From these figures, we can clearly observe that the RF model's predicted trajectories almost overlap with the actual anesthesia trajectories, indicating an exceptional predictive performance. This visual consistency underscores the RF model's superior accuracy and high stability in capturing the trends of actual trajectories.

\begin{table*}[t!]
\centering
\begin{threeparttable}
\caption{Comparison of Anesthesia Effects of Different VD-MADRL Methods with Human Experience}
\label{table_mape}
\begin{tabular}{|c|c|c|c|c|c|c|c|c|c|c|c|}
\hline 
\multirow{3}{*}{\textbf{\makecell{Online\\/Offline}}} &\multirow{3}{*}{\textbf{\makecell{Method\\(Baseline/\\RF+)}}} & \multicolumn{5}{c|}{\textbf{General surgery}}& \multicolumn{5}{c|}{\textbf{Thoracic surgery}}\\
\cline{3-12}
& & \multirow{2}{*}{\textbf{$CR\uparrow$}} & \multicolumn{4}{c|}{\textbf{$MDPE\downarrow$}}&\multirow{2}{*}{\textbf{$CR\uparrow$}} & \multicolumn{4}{c|}{\textbf{$MDPE\downarrow$}}\\
\cline{4-7}\cline{9-12}
& & 
&\textit{mean}&\textit{max}&\textit{min}&\textit{std}& &\textit{mean}&\textit{max}&\textit{min}&\textit{std}\\
\hline
\textbf{Baseline} &\textbf{{\makecell{Human\\experience}}}&140325 &0.2058 & 0.4440& 0.0740& 0.0687&91512 &0.2167 &0.4780 & 0.0660 & 0.0788 \\ 
\hline
\multirow{7}{*}{\textbf{Online}} &\textbf{VDN}&\textcolor{red}{163361} &\textcolor{red}{0.0864} & \textcolor{red}{0.3262}& \textcolor{red}{0.0046}& \textcolor{red}{0.0517}&96782 &0.2156 &0.2952 &0.0564 &0.0472  \\  
&\textbf{QTRAN}& 162044&0.0882 &0.2337 &0.0072 &0.0653 &96084 &0.2166 &0.2949 &0.0299 &0.0535  \\  
&\textbf{QPLEX}& \textcolor{blue}{157475}&\textcolor{blue}{0.1387} & \textcolor{blue}{0.2790} &\textcolor{blue}{0.0102} & \textcolor{blue}{0.0575}&\textcolor{red}{98762} &\textcolor{red}{0.1917} &\textcolor{red}{0.2810} &\textcolor{red}{0.0119} &\textcolor{red}{0.0665}  \\ 
&\textbf{Qmix}&162476 &0.0888 &0.2389 &0.0065 &0.0606 &\textcolor{blue}{95831} &\textcolor{blue}{0.2166} &\textcolor{blue}{0.3284} &\textcolor{blue}{0.0036} &\textcolor{blue}{0.0768}  \\ 
&\textbf{QATTEN}&162199 &0.1007 &0.2597 &0.0160 &0.0478 &98170 &0.1991 &0.2688 &0.0367 &0.0516  \\ 
&\textbf{OW\_QMIX}& 163199&0.0823 &0.2448 &0.0053 &0.0566 &96741 &0.2141 &0.3016 &0.0162 &0.0624  \\ 
&\textbf{CW\_QMIX}&160027 &0.1205 &0.2275 &0.0138 &0.0497 &96842 &0.2078 &0.3127 &0.0194 &0.0642  \\ 
\hline
\multirow{7}{*}{\textbf{Offline}} &\textbf{VDN}&158864 &0.1148 &0.3308 &0.0295 &0.0581 & 98831&0.2012 &0.3232 &0.0548 &0.0515  \\ 
&\textbf{QTRAN}&160492 &0.1151 &0.2290 &0.0226 &0.0494 & 97612&0.2057 &0.3174 &0.0270 &0.0652  \\ 
&\textbf{QPLEX}& \textcolor{red}{162375}&\textcolor{red}{0.0896} &\textcolor{red}{0.3272} &\textcolor{red}{0.0143} &\textcolor{red}{0.0640} & 96841&0.2145 &0.2866 &0.0199 &0.0400  \\ 
&\textbf{Qmix}&161831 &0.0963 &0.2340 &0.0214 &0.0472 &98428 &0.2027 &0.3079 &0.0318 &0.0594  \\ 
&\textbf{QATTEN}& 161313&0.1036 &0.2363 &0.0165 &0.0515 &97419 &0.2044 &0.2913 &0.0330 &0.0588  \\ 
&\textbf{OW\_QMIX}&158441 &0.1166 &0.3698 &0.0129 &0.0804 &\textcolor{blue}{96652} &\textcolor{blue}{0.2110} &\textcolor{blue}{0.3401} &\textcolor{blue}{0.0321} &\textcolor{blue}{0.0667}  \\ 
&\textbf{CW\_QMIX}&\textcolor{blue}{156951} &\textcolor{blue}{0.1437} &\textcolor{blue}{0.2347} &\textcolor{blue}{0.0326} &\textcolor{blue}{0.0510} &\textcolor{red}{100012} &\textcolor{red}{0.1883} &\textcolor{red}{0.3133} &\textcolor{red}{0.0121} &\textcolor{red}{0.0580}  \\ 
\hline
\end{tabular}
\begin{tablenotes}
\item[*] \textcolor{red}{Red text} is the best performing of different training modes in each dataset, \textcolor{blue}{blue text} is the worst performing of different training modes in each dataset.
\end{tablenotes}
\end{threeparttable}
\end{table*}

\begin{figure}[t!]
    \centering
    \begin{subfigure}[t]{0.45\textwidth}
        \includegraphics[width=\linewidth]{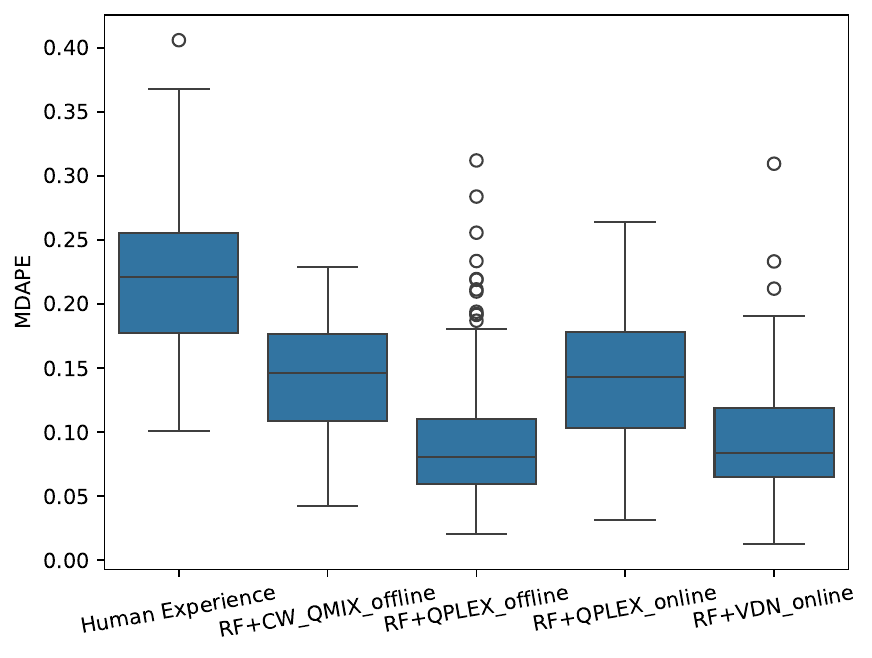}
        \caption{MDAPE Distribution for General Surgery Dataset}
        \label{fig:SG_mse}
    \end{subfigure}
    \hfill
    \begin{subfigure}[t]{0.45\textwidth}
        \includegraphics[width=\linewidth]{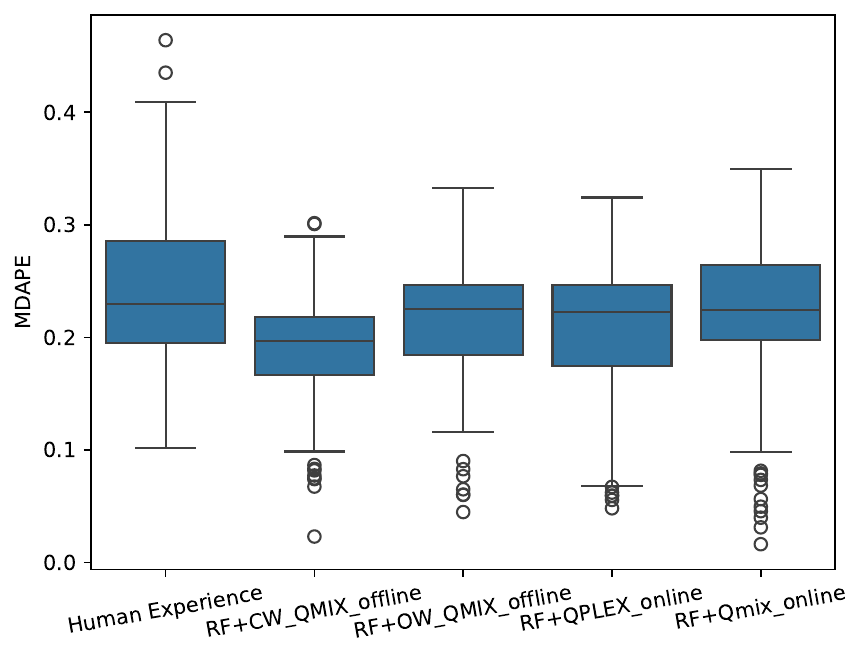}
        \caption{MDAPE Distribution for Thoracic Surgery Dataset}
        \label{fig:bis}
    \end{subfigure}
    \caption{Comparison of MDAPE for Selected Methods Against Human Experience in General and Thoracic Surgery Datasets.}
    \label{fig:box}
\end{figure}

\begin{figure*}[t!]
    \centering
    \begin{subfigure}[t]{0.23\textwidth}
        \includegraphics[width=\linewidth]{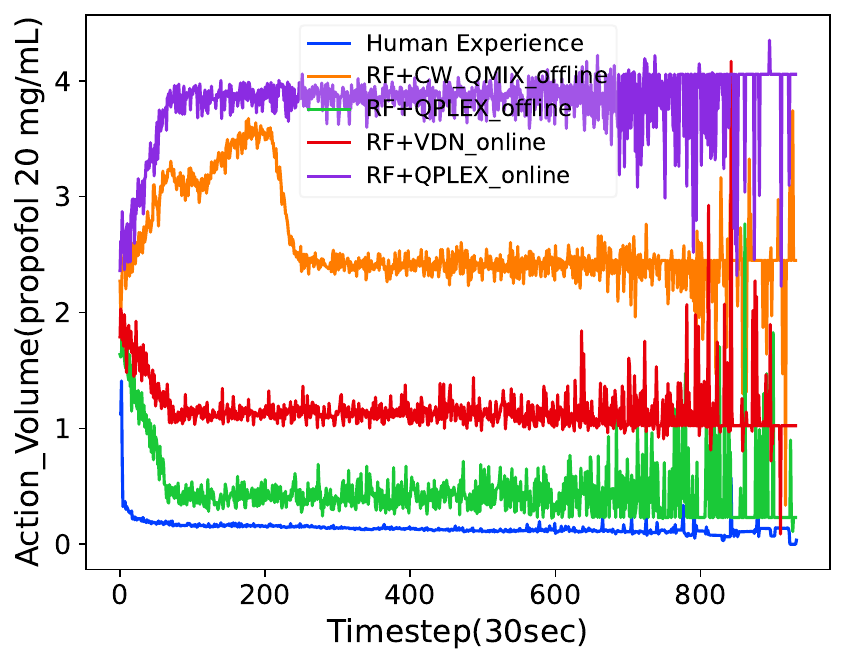}
        \caption{Action trajectories of PPF}
        \label{fig:SG_mse}
    \end{subfigure}
    \hfill
    \begin{subfigure}[t]{0.23\textwidth}
        \includegraphics[width=\linewidth]{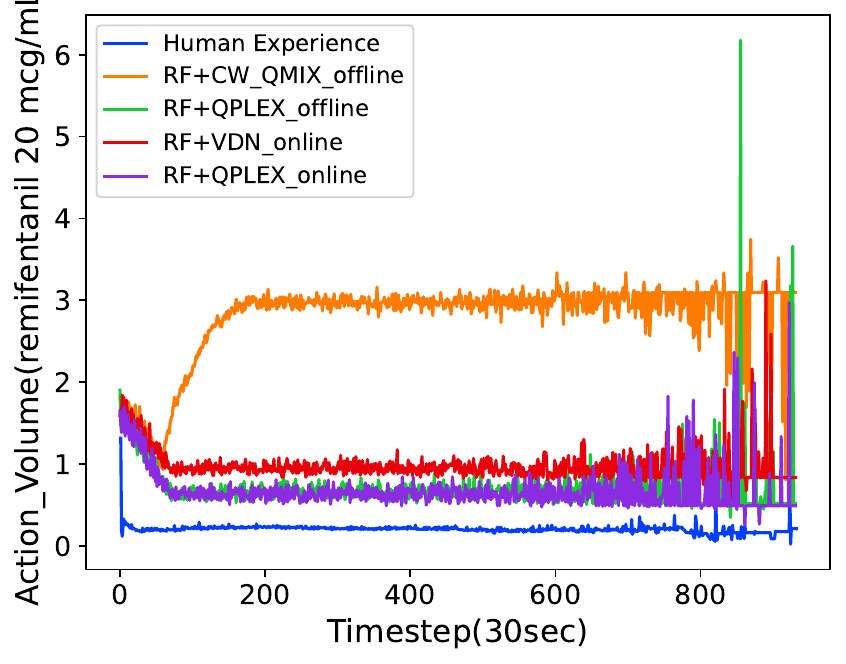}
        \caption{Action traj. of RFTN}
        \label{fig:bis}
    \end{subfigure}
    \hfill
    \begin{subfigure}[t]{0.23\textwidth}
        \includegraphics[width=\linewidth]{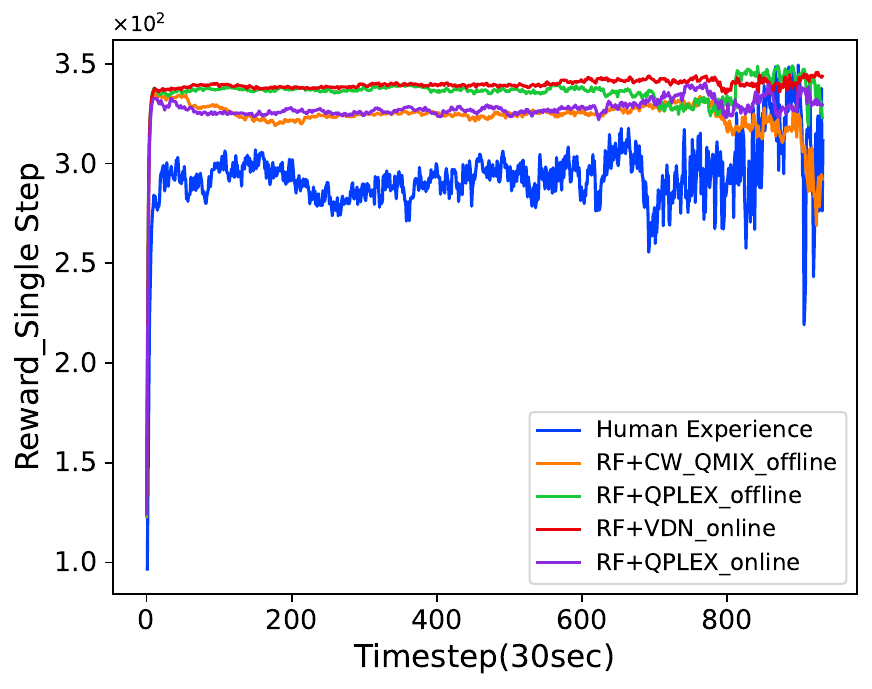}
        \caption{Single-step reward traj.}
        \label{fig:gru}
    \end{subfigure}
    \hfill
    \begin{subfigure}[t]{0.23\textwidth}
        \includegraphics[width=\linewidth]{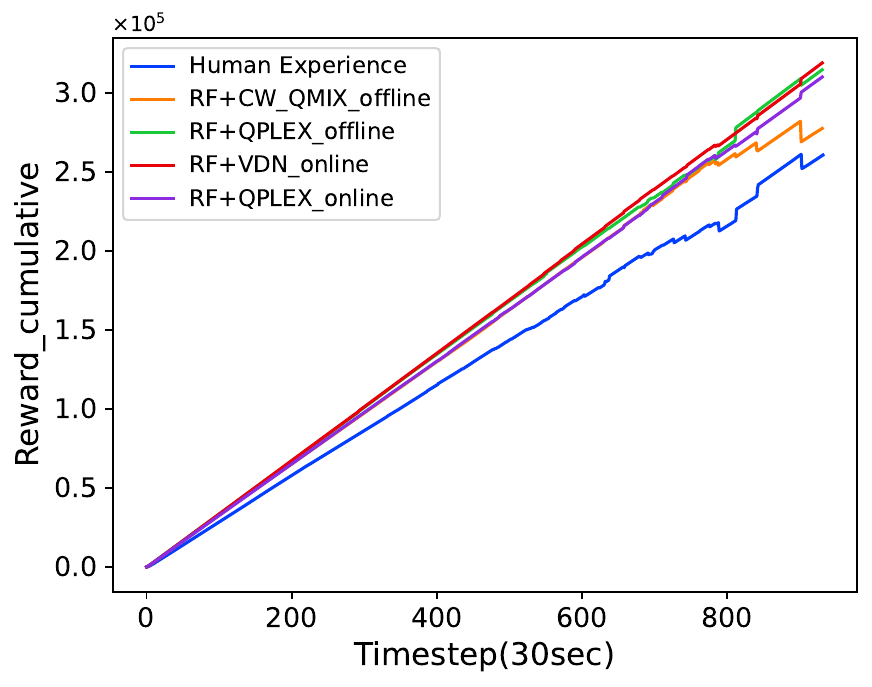}
        \caption{Cumulative reward traj.}
        \label{fig:rr}
    \end{subfigure}\\
    \begin{subfigure}[t]{0.23\textwidth}
        \includegraphics[width=\linewidth]{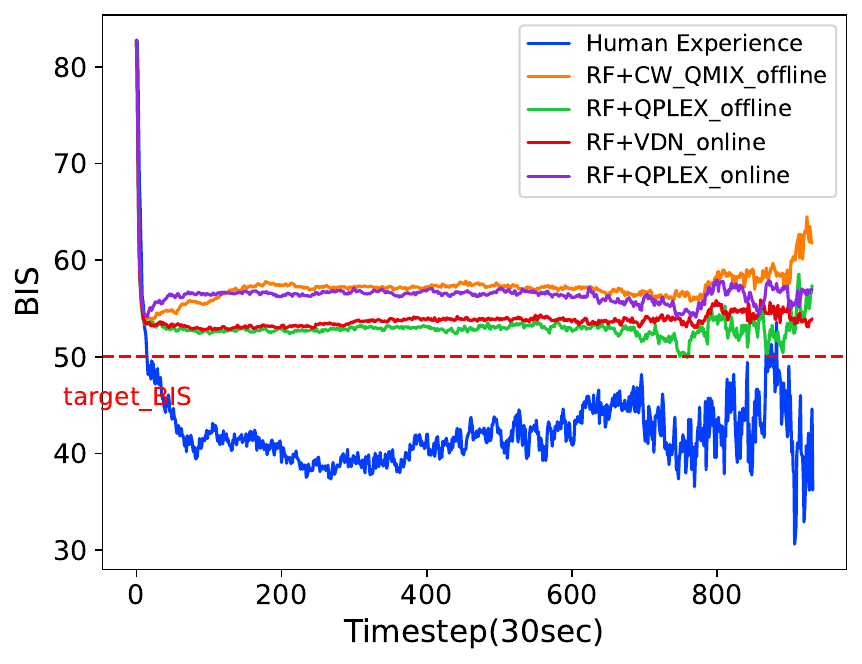}
        \caption{BIS trajectories}
        \label{fig:SG_mse}
    \end{subfigure}
    \hfill
    \begin{subfigure}[t]{0.23\textwidth}
        \includegraphics[width=\linewidth]{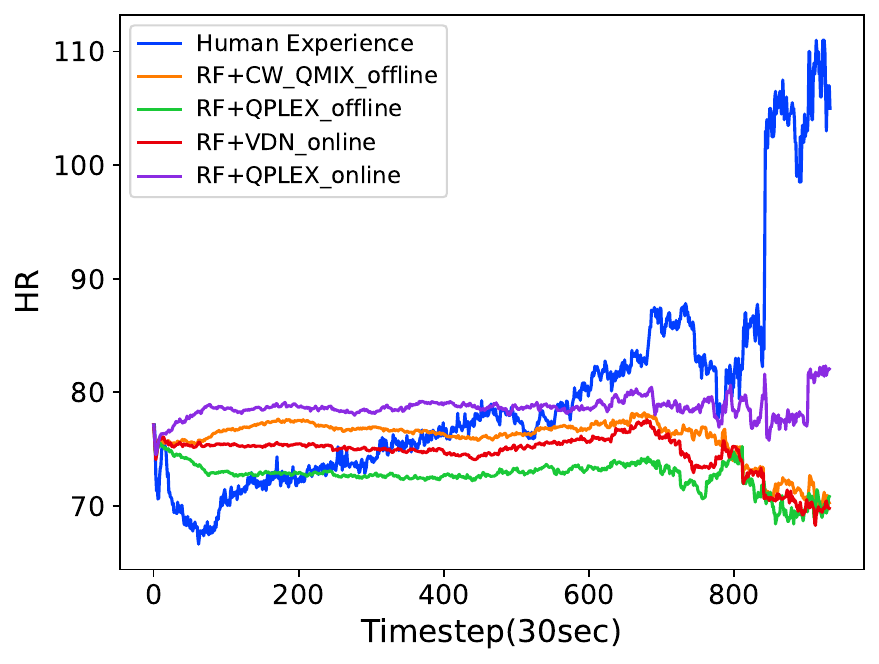}
        \caption{HR trajectories}
        \label{fig:bis}
    \end{subfigure}
    \hfill
    \begin{subfigure}[t]{0.23\textwidth}
        \includegraphics[width=\linewidth]{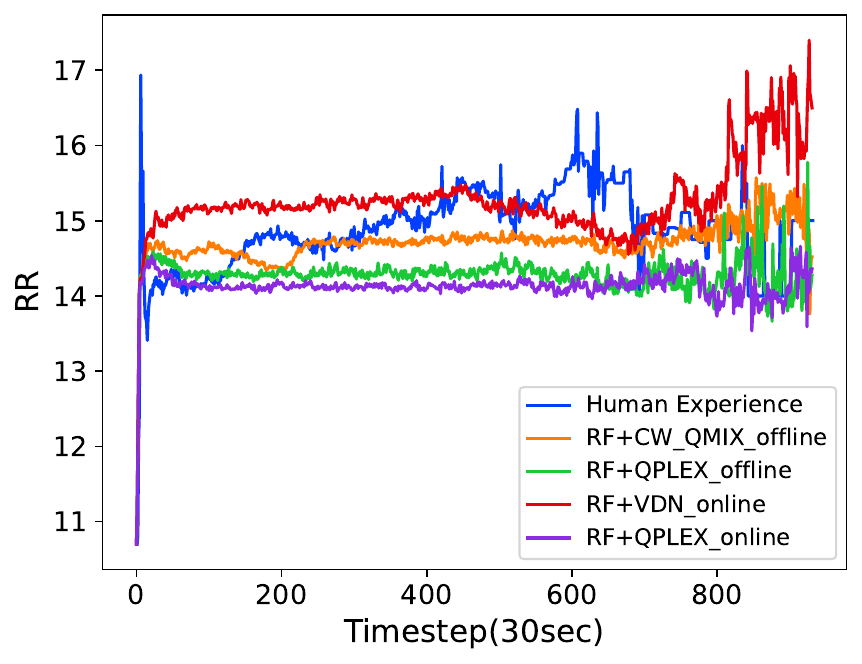}
        \caption{RR trajectories}
        \label{fig:gru}
    \end{subfigure}
    \hfill
    \begin{subfigure}[t]{0.23\textwidth}
        \includegraphics[width=\linewidth]{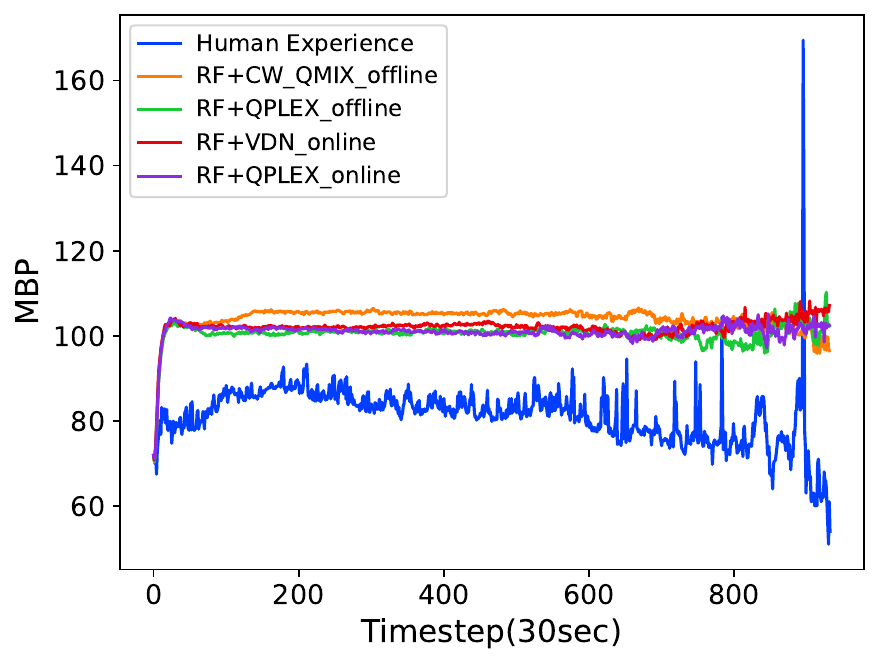}
        \caption{MBP trajectories}
        \label{fig:rr}
    \end{subfigure}\\
        \begin{subfigure}[t]{0.23\textwidth}
        \includegraphics[width=\linewidth]{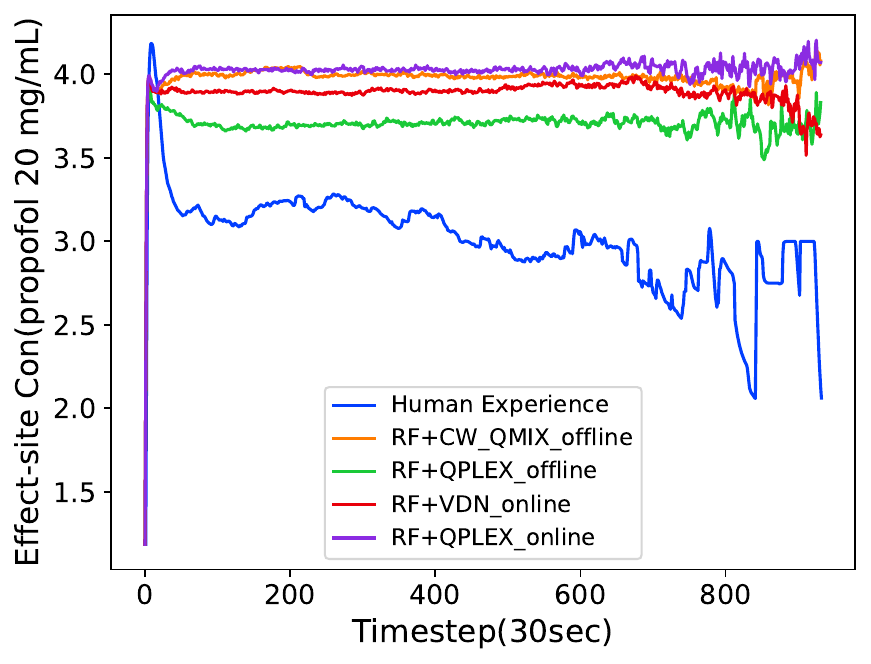}
        \caption{CE trajectories of PPF}
        \label{fig:SG_mse}
    \end{subfigure}
    \hfill
    \begin{subfigure}[t]{0.23\textwidth}
        \includegraphics[width=\linewidth]{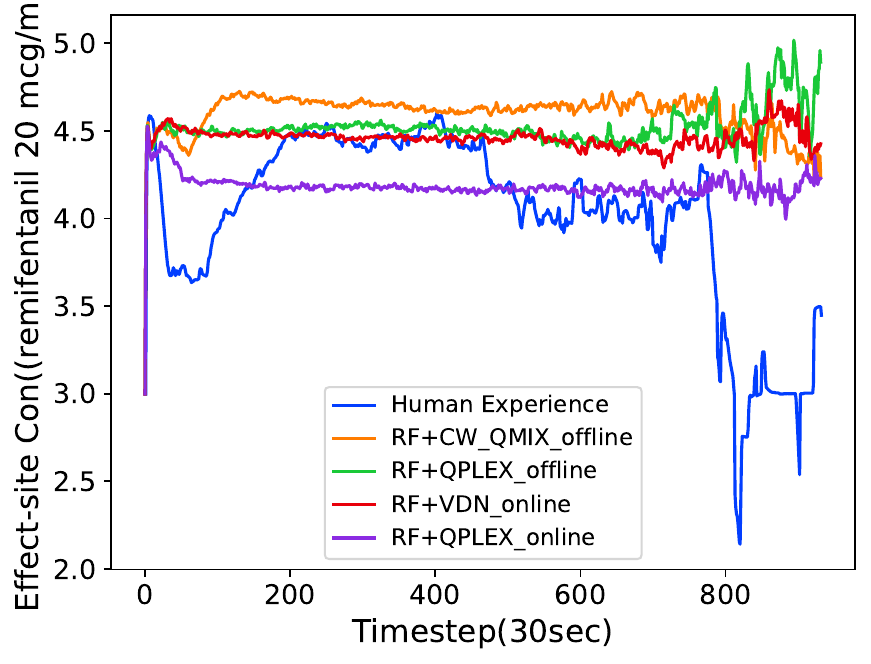}
        \caption{CE traj. of RFTN}
        \label{fig:bis}
    \end{subfigure}
    \hfill
    \begin{subfigure}[t]{0.23\textwidth}
        \includegraphics[width=\linewidth]{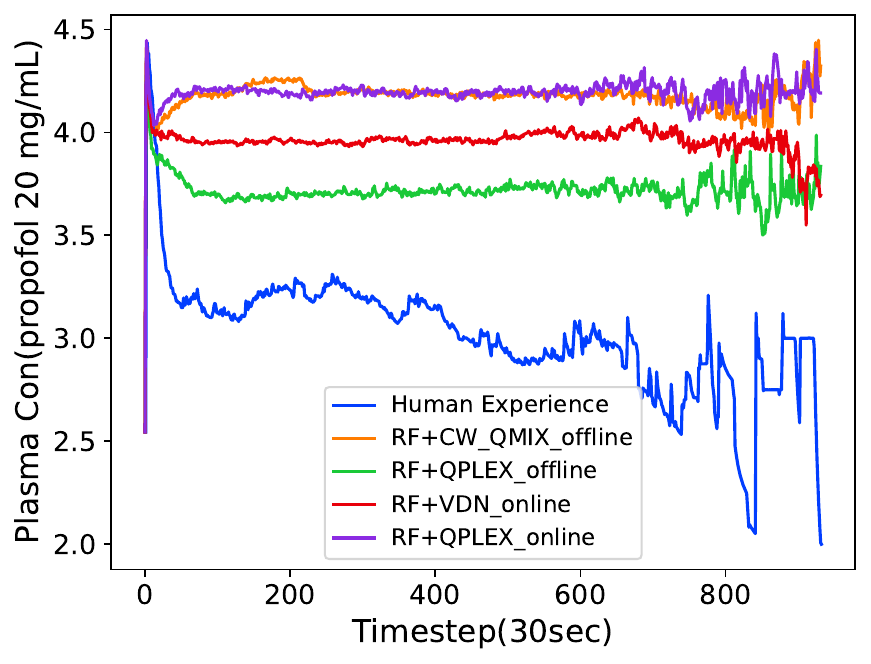}
        \caption{CP trajectories of PPF}
        \label{fig:gru}
    \end{subfigure}
    \hfill
    \begin{subfigure}[t]{0.23\textwidth}
        \includegraphics[width=\linewidth]{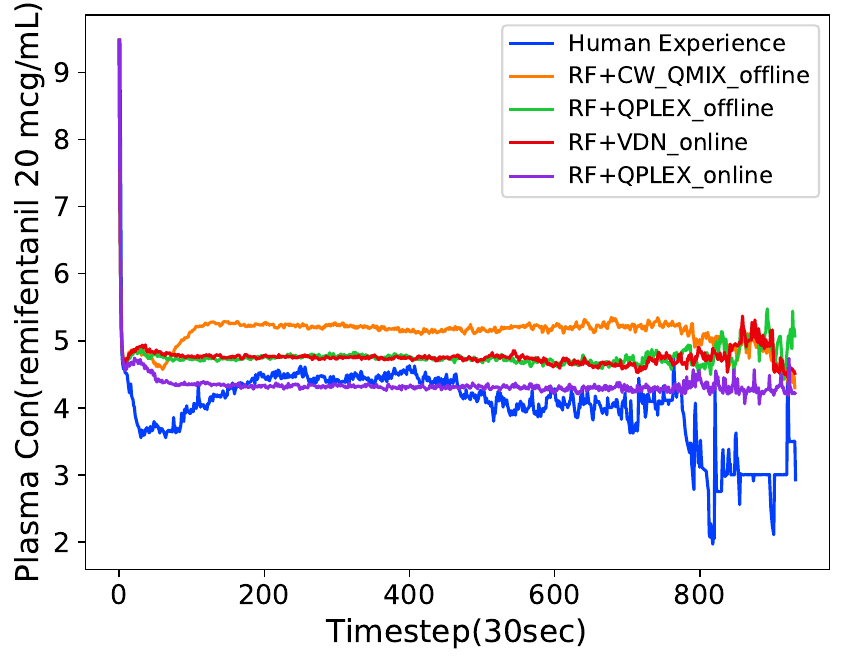}
        \caption{CP trajectories of RFTN}
        \label{fig:rr}
    \end{subfigure}
    \caption{In General Surgery Dataset: Comparison of Traj. Generated by VD-MADRL and Human Experience.}
    \label{fig:rl_general}
\end{figure*}

\begin{figure*}[t!]
    \centering
    \begin{subfigure}[t]{0.23\textwidth}
        \includegraphics[width=\linewidth]{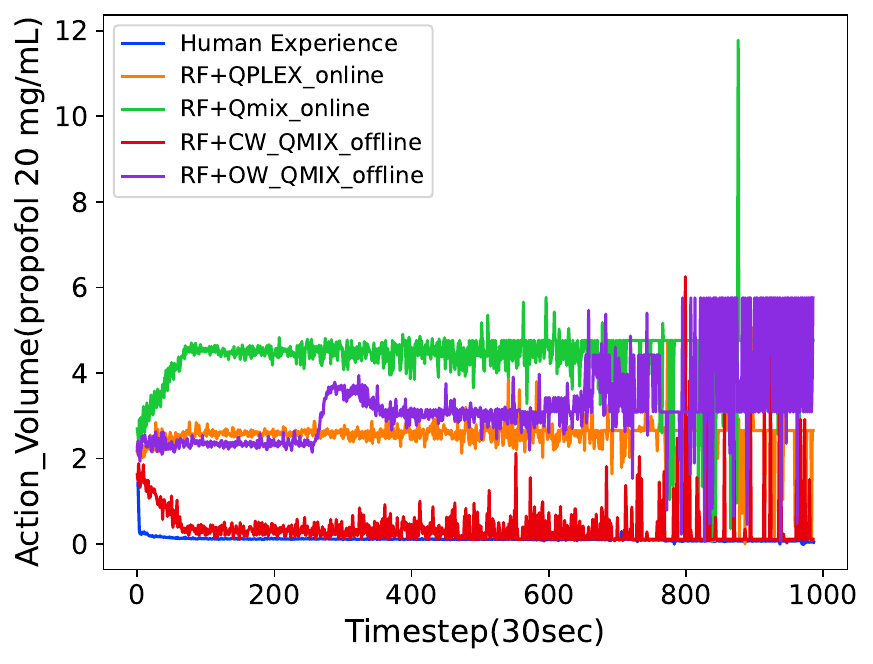}
        \caption{Action trajectories of PPF}
    \end{subfigure}
    \hfill
    \begin{subfigure}[t]{0.23\textwidth}
        \includegraphics[width=\linewidth]{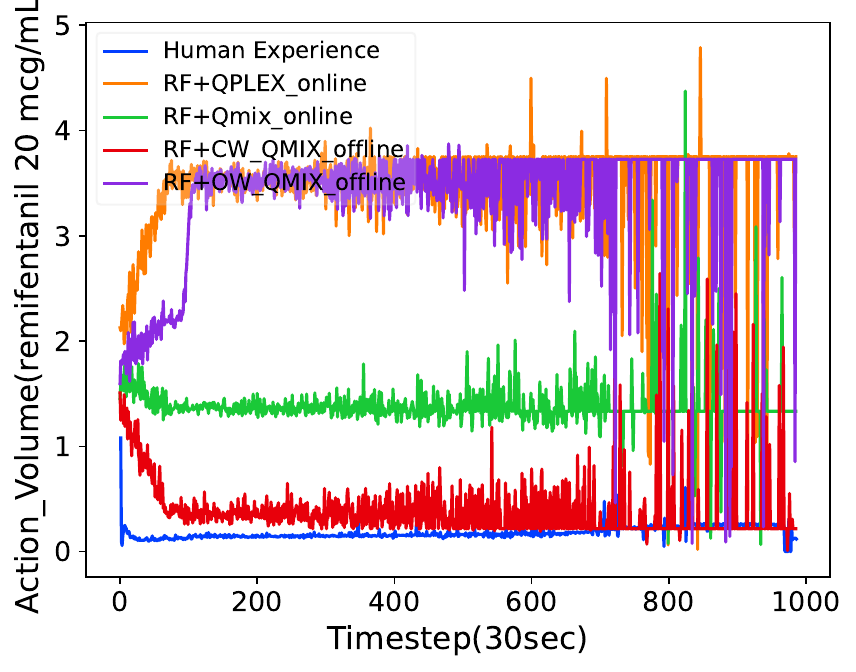}
        \caption{Action traj. of RFTN}
    \end{subfigure}
    \hfill
    \begin{subfigure}[t]{0.23\textwidth}
        \includegraphics[width=\linewidth]{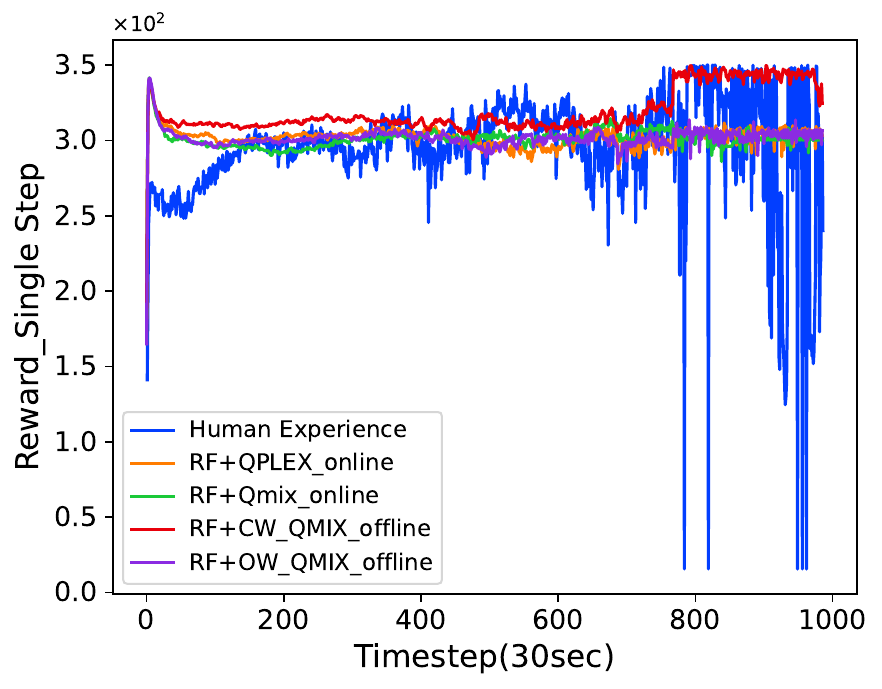}
        \caption{Single-step reward traj.}
    \end{subfigure}
    \hfill
    \begin{subfigure}[t]{0.23\textwidth}
        \includegraphics[width=\linewidth]{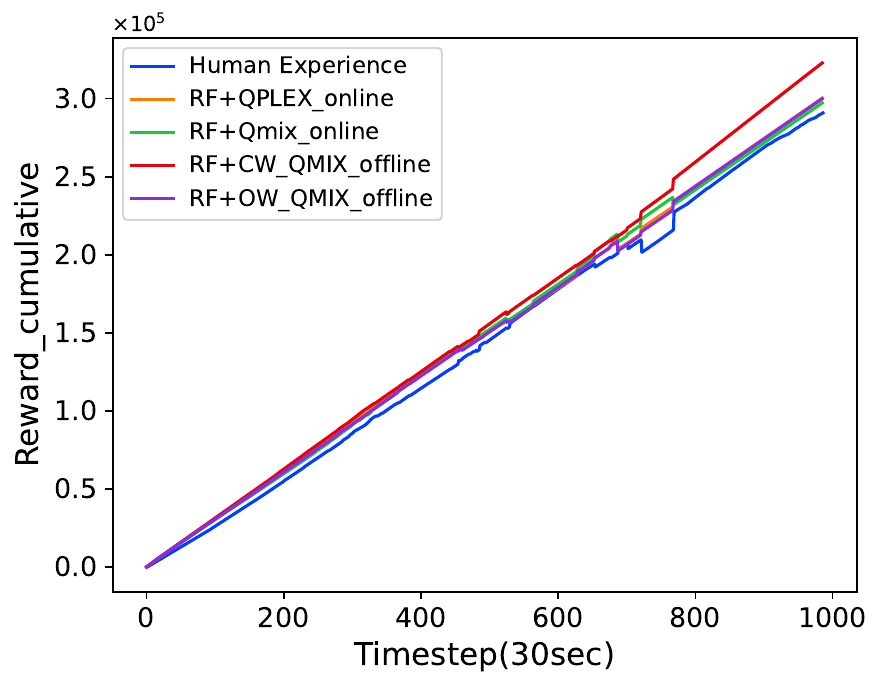}
        \caption{Cumulative reward traj.}
    \end{subfigure}\\
    \begin{subfigure}[t]{0.23\textwidth}
        \includegraphics[width=\linewidth]{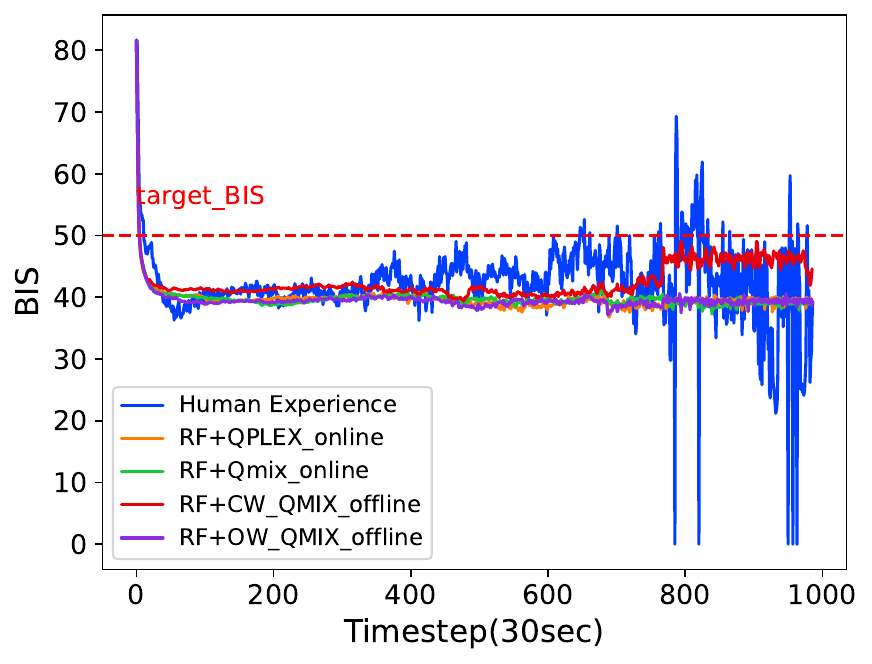}
        \caption{BIS trajectories}
    \end{subfigure}
    \hfill
    \begin{subfigure}[t]{0.23\textwidth}
        \includegraphics[width=\linewidth]{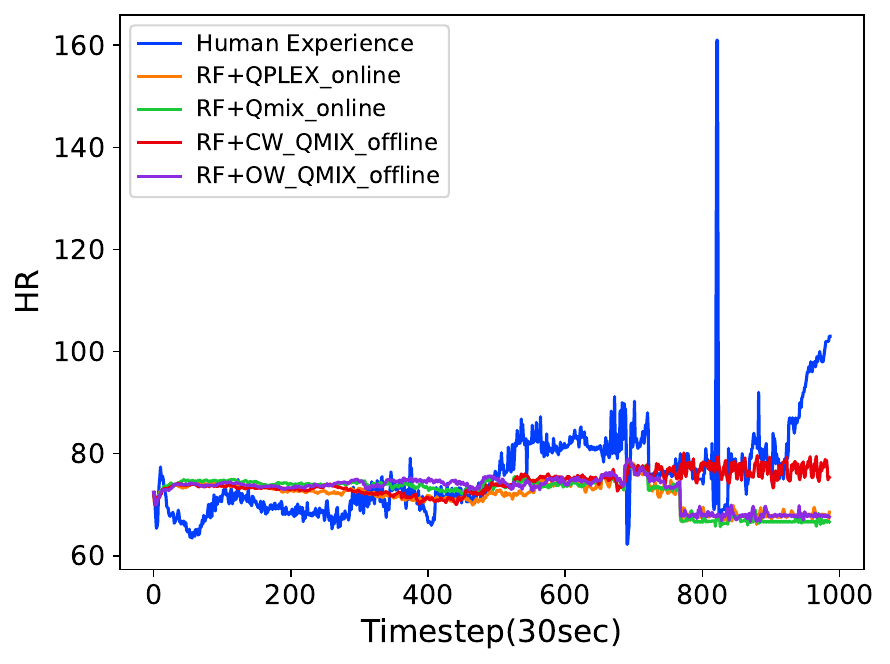}
        \caption{HR trajectories}
    \end{subfigure}
    \hfill
    \begin{subfigure}[t]{0.23\textwidth}
        \includegraphics[width=\linewidth]{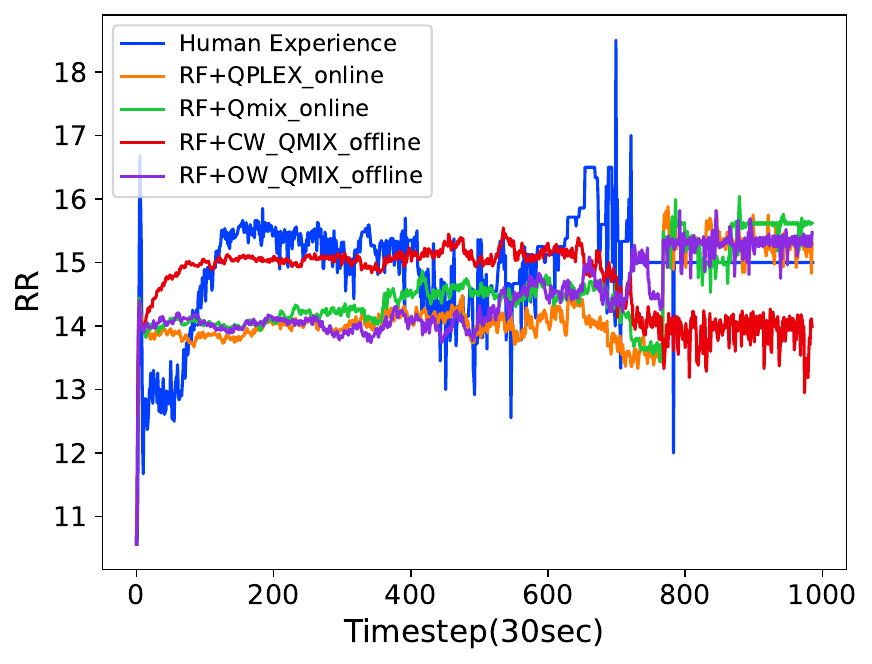}
        \caption{RR trajectories}
    \end{subfigure}
    \hfill
    \begin{subfigure}[t]{0.23\textwidth}
        \includegraphics[width=\linewidth]{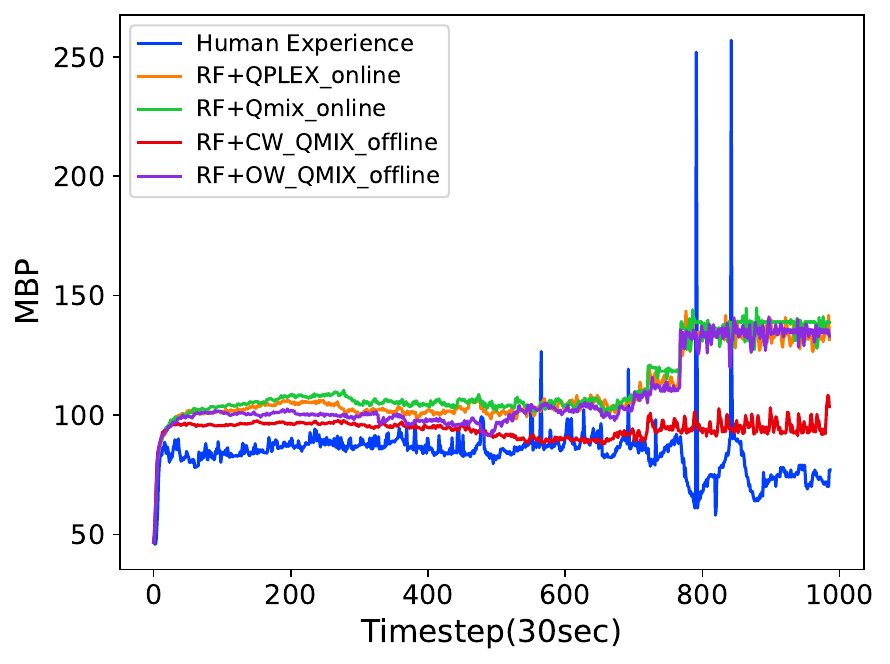}
        \caption{MBP trajectories}
    \end{subfigure}\\
        \begin{subfigure}[t]{0.23\textwidth}
        \includegraphics[width=\linewidth]{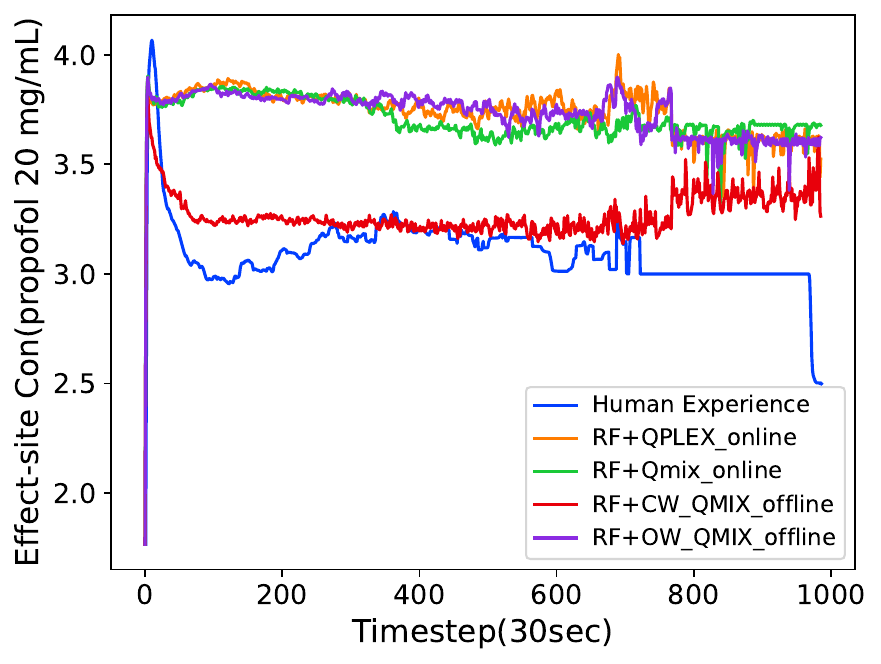}
        \caption{CE trajectories of PPF}
    \end{subfigure}
    \hfill
    \begin{subfigure}[t]{0.23\textwidth}
        \includegraphics[width=\linewidth]{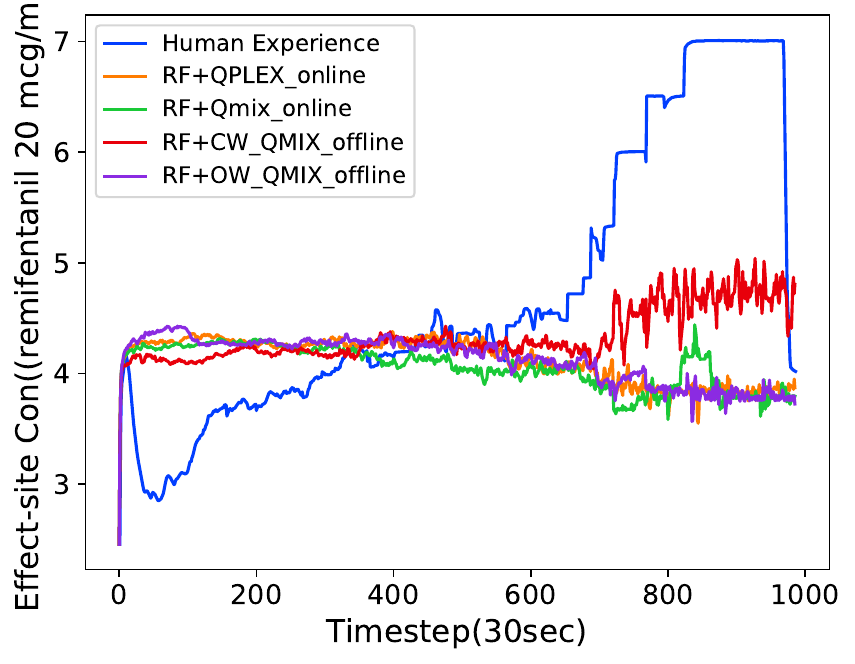}
        \caption{CE trajectories of RFTN}
    \end{subfigure}
    \hfill
    \begin{subfigure}[t]{0.23\textwidth}
        \includegraphics[width=\linewidth]{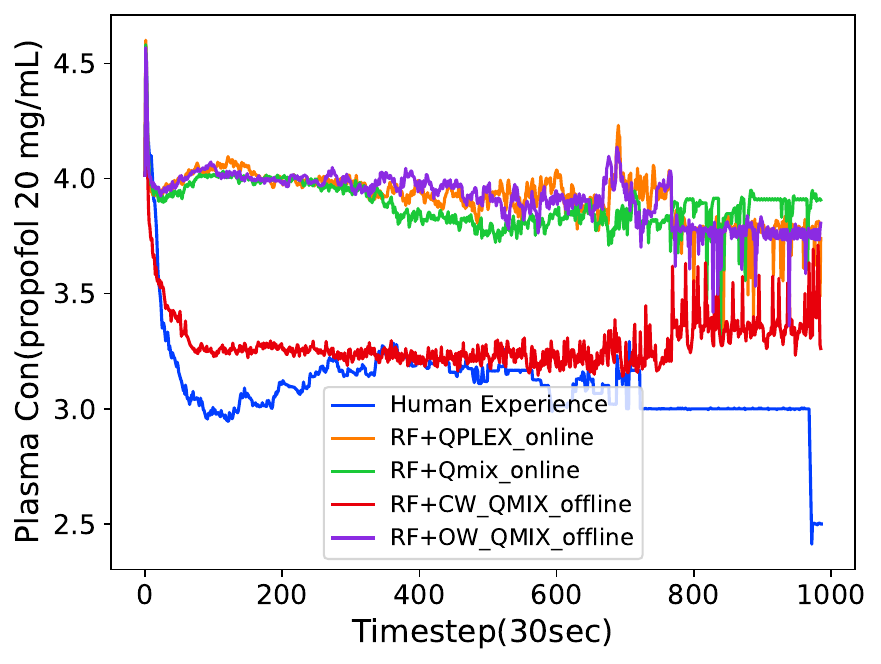}
        \caption{CP trajectories of PPF}
    \end{subfigure}
    \hfill
    \begin{subfigure}[t]{0.23\textwidth}
        \includegraphics[width=\linewidth]{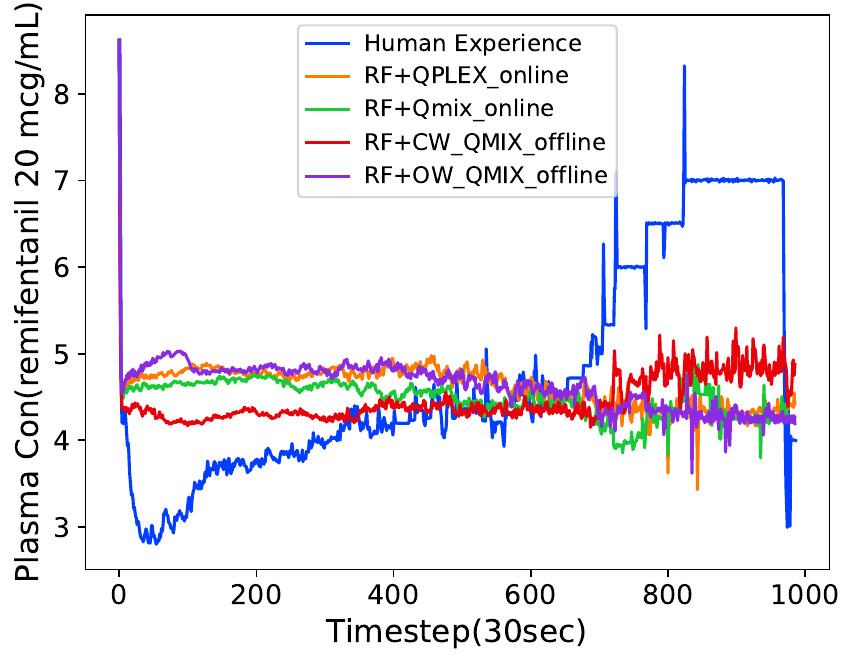}
        \caption{CP trajectories of RFTN}
    \end{subfigure}
    \caption{In Thoracic Surgery Dataset: Comparison of Traj. Generated by VD-MADRL and Human Experience.}
    \label{fig:rl_thoracic}
\end{figure*}

\subsection{Anesthesia Effect Analysis of VD-MADRL}
Table~\ref{table_mape} presents a comprehensive comparison of the performance of seven state-of-the-art value decomposition models (VDN, QMIX, CW\_QMIX, OW\_QMIX, QPLEX, QTRAN, and Qatten) against human experience across two datasets: General Surgery and Thoracic Surgery. The evaluation metrics include the mean, maximum, minimum, and standard deviation of MDPE and the average CR across all test cases. This assessment is conducted under both online and offline training modes.

The analysis further shows that even the worst performing model outperforms human experience in both datasets, highlighting the potential of VD-MADRL methods in enhancing collaborative management of multiple anesthetics. The differences in model performance across datasets and training modes highlight the importance of selecting appropriate value decomposition methods and training strategies for specific clinical scenarios.


Fig.~\ref{fig:box} provides a detailed comparative analysis of the Median Absolute Deviation of Performance Error (MDAPE)~\cite{mdpemdape} in BIS deviation from the target BIS value for both the best and worst-performing VD-MADRL models and human experience across two datasets: General Surgery and Thoracic Surgery. The x-axis represents selected VD-MADRL models and human experience, while the y-axis denotes the MDAPE values. Each box plot illustrates the distribution of MDAPE values, where the central box captures the interquartile range (IQR), the median line indicates the central tendency. The lower median line and narrower boxes in the box plots signify that the models' anesthesia effect are closer to the target BIS values and have less variability. 

Specifically, in both general surgery dataset and the thoracic surgery dataset, human experience shows a relatively high median MDAPE with a wide IQR, indicating substantial variability in performance. The RF+CW\_QMIX\_offline model exhibits a lower median MDAPE and a narrower IQR, reflecting more consistent performance with smaller deviations from the target BIS values. This trend continues with the other models. Notably, the best performing model RF+VDN\_online model in general surgery dataset and RF+CW\_QMIX\_offline model in thoracic surgery dataset also achieves the lowest median MDAPE and the smallest IQR, indicating superior and consistent performance with minimal deviation from the target BIS values. 

Overall, our VD-MADRL models consistently exhibit lower median MDAPE and narrower IQR compared to human experience, indicating superior and more stable performance. Even though some models present outliers, these deviations are generally lower than those observed in human experience. This suggests that our approach not only enhances the accuracy of anesthesia management but also reduces the variability and incidence of extreme errors, further demonstrating the robustness and reliability of VD-MADRL methods in clinical applications.

Fig.~\ref{fig:rl_general} and Fig.~\ref{fig:rl_thoracic} show the four models with the best and worst performance under different training modes in the general surgery dataset and thoracic surgery dataset, respectively, for comparison with human experience. The X-axis represents the time step of the entire anesthesia process, with each time step interval of 30 seconds; the Y-axis represents the average value of each trajectory in the test set.
\begin{itemize}
    \item Anesthetic Infusion Trajectories (a, b). sub-figures (a) and (b) show the infusion doses of propofol (PPF) and remifentanil (RFTN) at each time step for general surgery (Fig.~\ref{fig:rl_general}) and thoracic surgery (Fig.~\ref{fig:rl_thoracic}), respectively. Human experience typically administer a large initial dose followed by a near-constant rate, exhibiting minimal fluctuation over time. In contrast, our models demonstrate rapid and variable dosing at each time step, reflecting higher flexibility. Notably, the best-performing models administer significantly lower doses compared to the worst-performing ones, indicating potential areas for further optimization.
    \item Reward Trajectories (c, d). sub-figures (c) and (d) in both datasets display the single-step and cumulative rewards. Our models consistently achieve higher rewards than human experience, suggesting superior performance in terms of the reward mechanism.
    \item Anesthesia State Indicators Trajectories (e, f, g, h). sub-figures (e) through (h) illustrate the trajectories of BIS, heart rate (HR), respiratory rate (RR), and mean arterial blood pressure (MBP). Our models outperform human experience by maintaining these metrics closer to target values with greater stability. Human experience, particularly in the later stages of anesthesia, exhibit significant fluctuations, occasionally reaching hazardous levels (e.g., HR and BP exceeding safe limits).
    \item Drug Concentration Trajectories (i, j, k, l). sub-figures (i) through (l) show the blood plasma concentrations and effect-site concentrations of Propofol and Remifentanil. Again, our models manage these indicators with greater precision and stability compared to human experience.
\end{itemize}
In summary, our models exhibit exceptional flexibility and precision in multiple anesthetics collaborative control, significantly enhancing anesthesia effect. They outperform human experience in reward mechanisms and anesthesia state control, particularly by maintaining more stable trajectories in the later stages of anesthesia. However, the variation in infusion doses between the best and worst-performing models indicates room for further refinement. 

\begin{figure*}[t!]
    \centering
    \begin{subfigure}[t]{0.23\textwidth}
        \includegraphics[width=\linewidth]{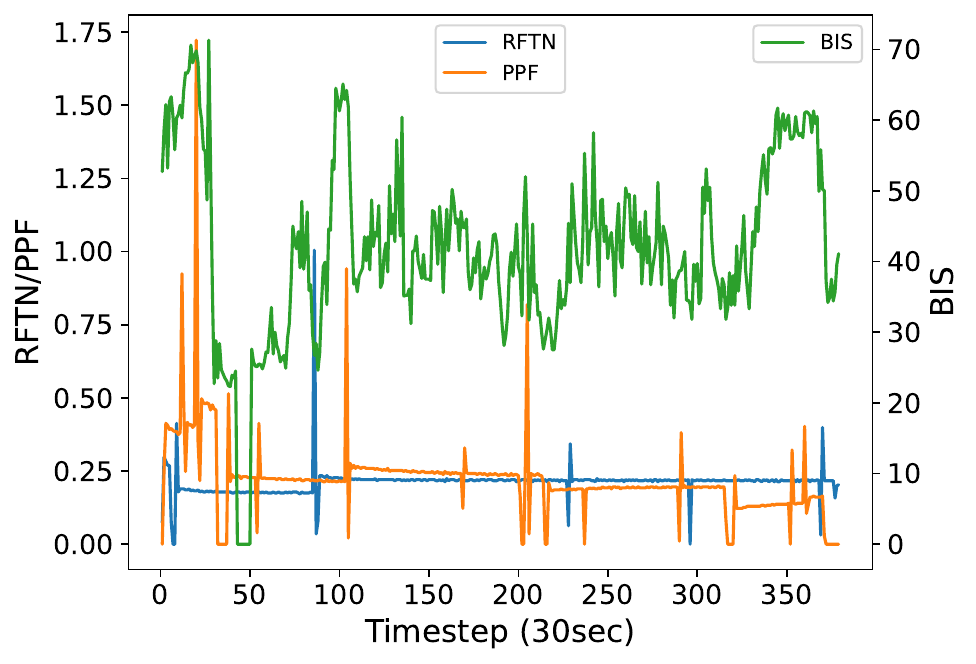}
        \caption{Human Experience\\(vs. RF+VDN\_online)}
        \label{fig:SG_mse}
    \end{subfigure}
    \hfill
    \begin{subfigure}[t]{0.23\textwidth}
        \includegraphics[width=\linewidth]{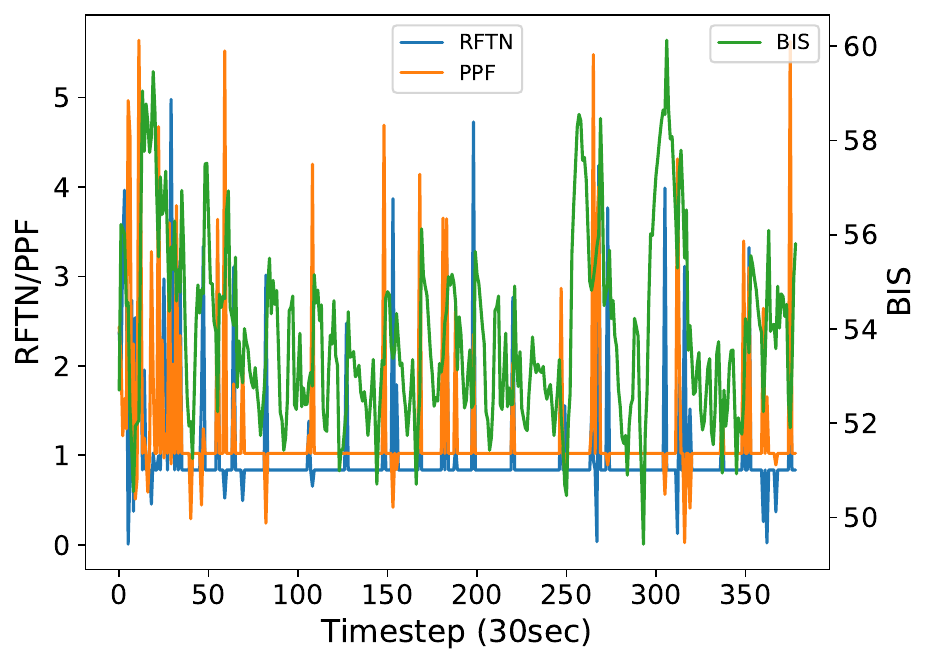}
        \caption{By RF+VDN\_online}
        \label{fig:bis}
    \end{subfigure}
    \hfill
    \begin{subfigure}[t]{0.23\textwidth}
        \includegraphics[width=\linewidth]{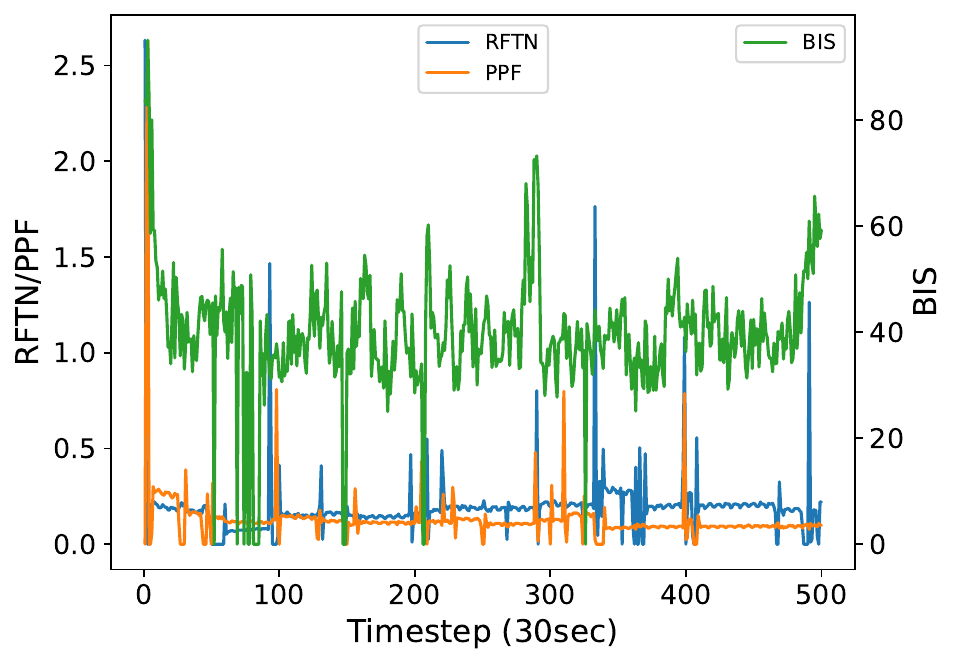}
        \caption{Human Experience\\(vs. RF+QPLEX\_offline)}
        \label{fig:SG_mse}
    \end{subfigure}
    \hfill
    \begin{subfigure}[t]{0.23\textwidth}
        \includegraphics[width=\linewidth]{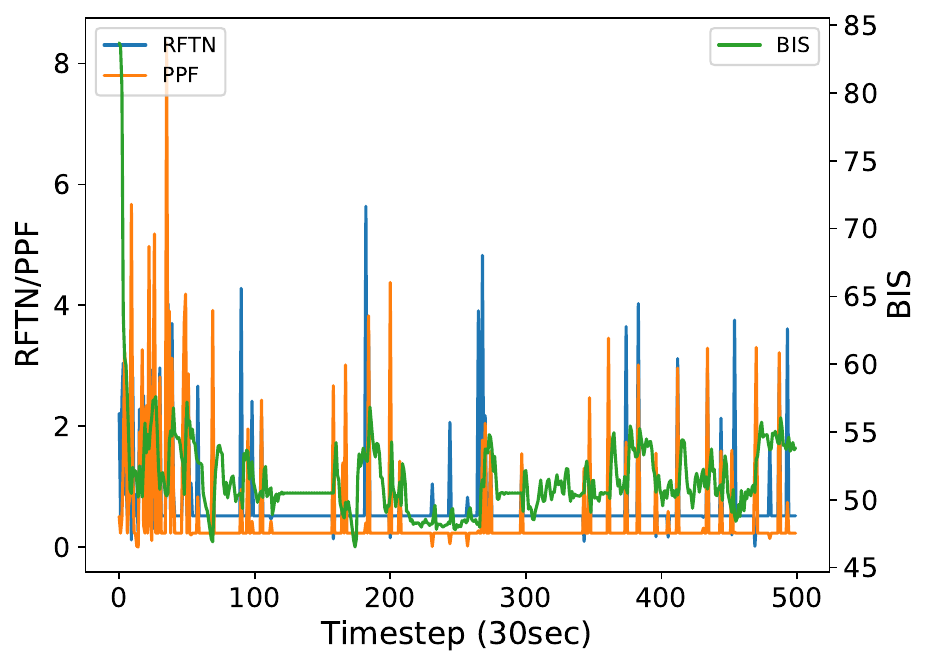}
        \caption{By RF+QPLEX\_offline}
        \label{fig:bis}
    \end{subfigure}
    \caption{Comparative Analysis of Anesthesia Trajectories in General Surgery Dataset: Human Experience vs. Ours.}
    \label{fig:case_general} 
    \begin{subfigure}[t]{0.23\textwidth}
        \includegraphics[width=\linewidth]{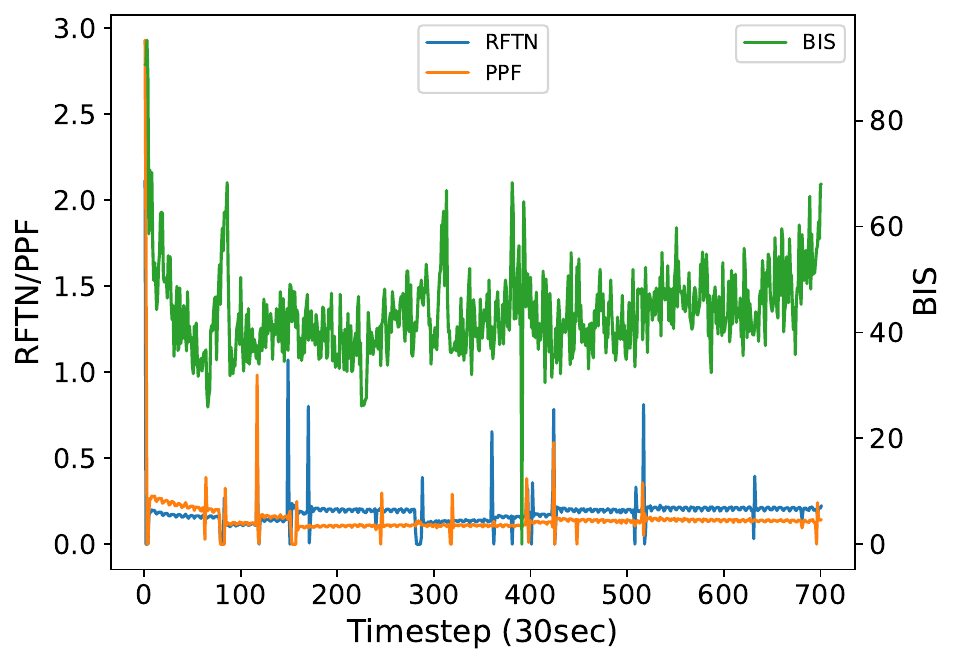}
        \caption{Human Experience\\(vs. RF+QPLEX\_online)}
        \label{fig:SG_mse}
    \end{subfigure}
    \hfill
    \begin{subfigure}[t]{0.23\textwidth}
        \includegraphics[width=\linewidth]{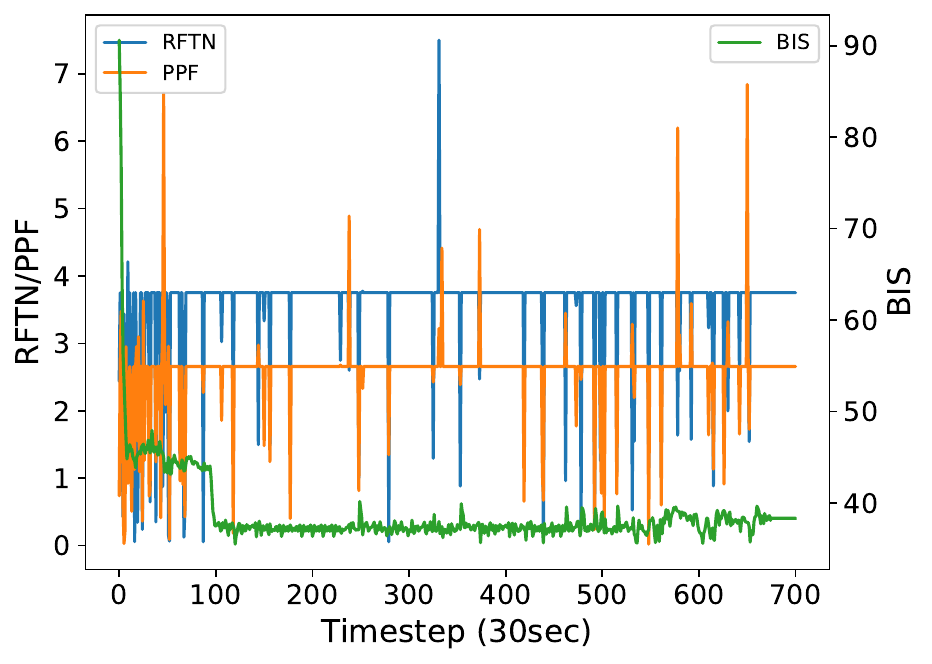}
        \caption{By RF+QPLEX\_online}
        \label{fig:bis}
    \end{subfigure}
    \hfill
    \begin{subfigure}[t]{0.23\textwidth}
        \includegraphics[width=\linewidth]{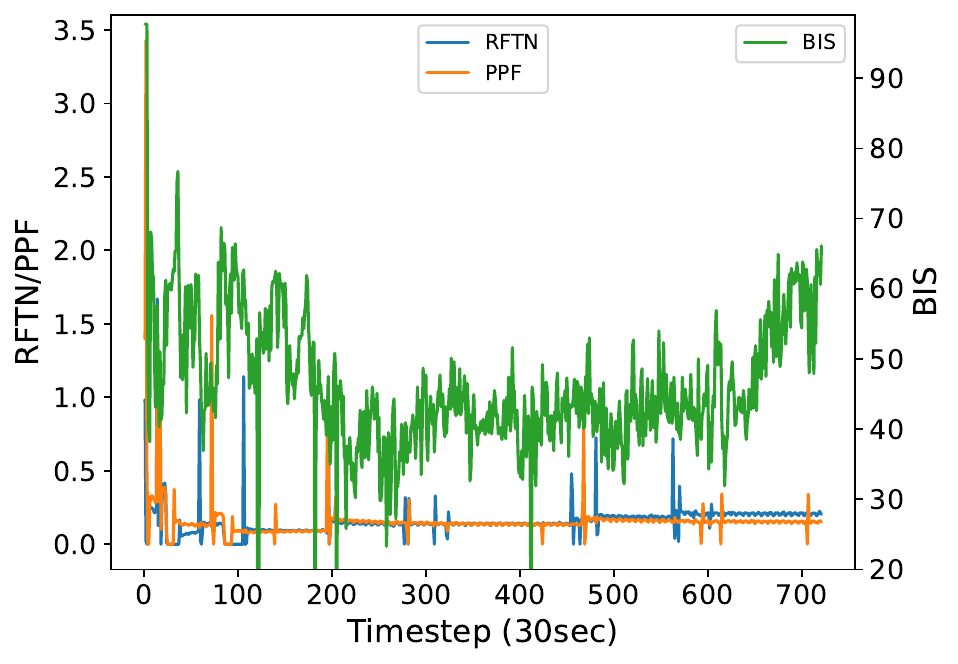}
        \caption{Human Experience\\(vs. RF+CW\_QMIX\_offline)}
        \label{fig:SG_mse}
    \end{subfigure}
    \hfill
    \begin{subfigure}[t]{0.24\textwidth}
        \includegraphics[width=\linewidth]{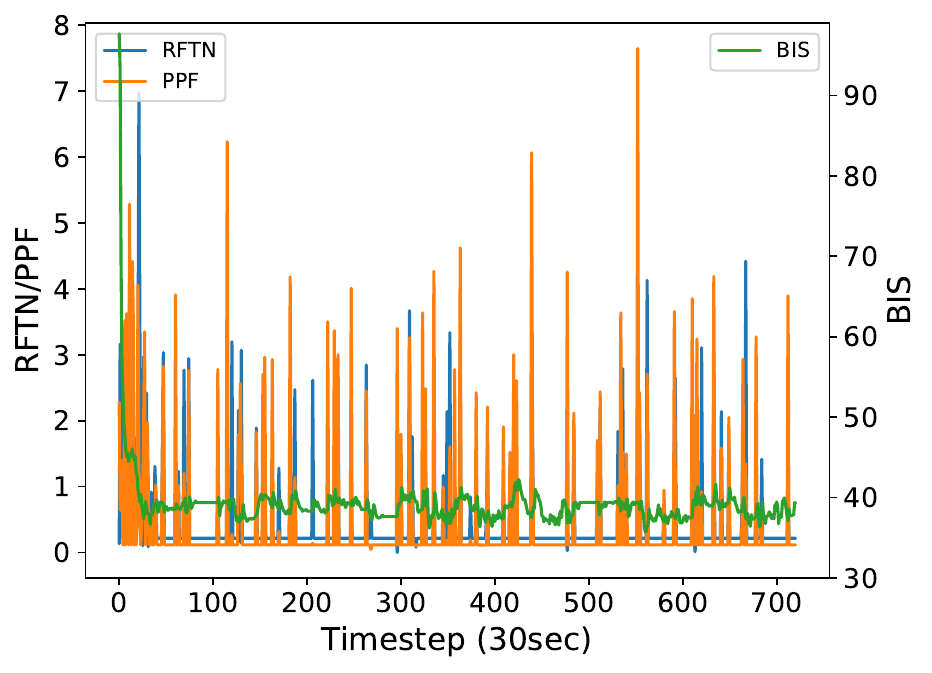}
        \caption{By RF+CW\_QMIX\_offline}
        \label{fig:bis}
    \end{subfigure}
    \caption{Comparative Analysis of Anesthesia Trajectories in Thoracic Surgery Dataset: Human Experience vs. Ours.}
    \label{fig:case_Thoracic}
\end{figure*}

\subsection{Case Study}
In Fig.~\ref{fig:case_general} and Fig.~\ref{fig:case_Thoracic}, we present a detailed comparison of anesthesia trajectories generated by human experience and our models under different training modes in general surgery and thoracic surgery datasets, respectively. We selected the best-performing models in both online and offline training modes for case analysis. The trajectories of two anesthetics and BIS produced by human experience are compared with those generated by our models, starting from the same initial states values and diverging as per the respective control strategies. 

Sub-figures (a) and (b) in both Fig.~\ref{fig:case_general} and Fig.~\ref{fig:case_Thoracic} compare human experience (with the same initial states values) and our model in the online mode. The human experience shows a pattern where anesthetics administration start with a large initial dose, followed by an almost constant rate with occasional adjustments. In contrast, within the first 60 time steps (approximately the initial one hour), our models make rapid and frequent adjustments to both Propofol and Remifentanil doses. This intensive modulation ensures that BIS reaches the target range quickly and efficiently. Subsequently, the model adjusts doses rapidly in response to BIS fluctuations, maintaining stability within the target level. 

sub-figures (c) and (d) in both Fig.~\ref{fig:case_general} and Fig.~\ref{fig:case_Thoracic} compare human experience (with the same initial states values) and our model in the offline mode. The human experience still begins with a substantial initial dose and then transitions to a near-constant rate with occasional changes. Our offline models in sub-figures (d) in both datasets, also show intensive dose adjustments within the first 60 time steps, achieving the target BIS range efficiently. The models continue to adjust doses dynamically to stabilize BIS within the desired range. In sub-figure (d), we can more easily see that as the BIS value fluctuates, the model continues to adjust the dose quickly. Once the BIS value stabilizes within the target range, the doses of the two anesthetics also become more stable. The model responds quickly to these fluctuations and adjusts the doses of the two anesthetics in a collaborative manner.

In summary, our models demonstrate superior flexibility and precision in multiple anesthetics collaborative control, significantly improving anesthesia effects. They outperform human experience by more rapidly adjusting doses to keep BIS stable within the target range. The close collaboration of two anesthetics doses with BIS changes, especially in sub-figure (d), highlights the model's ability to coordinately manage anesthetic administration. This fast and dynamically responsive approach highlights the potential of our model for clinical applications, ensuring more effective anesthesia management.

\subsection{Discussion}

Our VD-MADRL framework consistently outperforms human expertise in maintaining the stability and precision of anesthesia depth. However, certain limitations remain, particularly regarding the impact of high-dimensional state-action spaces on model performance and the generality of anesthesia control strategies.\textbf{Impact of high-dimensional state-action spaces:}
As more anesthesia state indicators or control measures are incorporated to enhance decision-making precision and achieve higher cumulative rewards, additional dimensions may be introduced into the state-action space. However, as the dimensions of the state-action space grow exponentially, our method faces the challenge of the "curse of dimensionality." This issue imposes greater demands on computational efficiency, memory requirements, and resource consumption. Therefore, one potential future direction is to improve sample efficiency or develop new methods that approximate optimal solutions for large state-action spaces.\textbf{Limitations of anesthetic dosing strategies:}
Table~\ref{table_mape} reveals significant performance differences for the same value-decomposition methods across different training modes and datasets. For instance, the QPLEX method performs worst in the online mode for the General Surgery dataset but achieves the best results in offline mode, whereas it exhibits the opposite behavior in the Thoracic Surgery dataset. Similarly, CW\_QMIX performs worst in the offline mode for the General Surgery dataset but excels under the same conditions in the Thoracic Surgery dataset. These results suggest that the effectiveness of each value-decomposition method may be highly environment-dependent, highlighting the need for careful selection of decomposition techniques tailored to specific clinical scenarios. The performance discrepancies across different datasets and training modes warrant further investigation to understand how varying surgical environments and training modes influence optimal anesthesia control strategies. Future work should focus on developing more robust methods capable of generalizing across diverse clinical environments.
In conclusion, while our VD-MADRL framework demonstrates significant advantages over human expertise, further research is needed to address its limitations in strategy generalization across different clinical scenarios and to improve its efficiency for high-dimensional real-time applications.

\section{Conclusion}
we propose an innovative framework VD-MADRL for clinical TIVA in closed-loop system. Our method effectively resolves the credit allocation problem among multiple anesthetics. It considers the simultaneous contributions of both anesthetic doses to the overall anesthesia effect through collaborative work. By integrating demographic data, DoA indicator BIS, vital signs, PK/PD data, and infused doses to design a multivariate  environment model. Our environment model better reflects individual differences and provides personalized anesthesia control. Our extensive experiments demonstrate our VD-MADRL framework offers finer granularity in dose adjustments and maintaining multiple anesthesia state indicators more stably at target levels compared to human experience, potentially enhancing patient safety and anesthesia quality.


\end{document}